\documentclass{aa}  

\usepackage{graphicx}
\usepackage{txfonts}
\usepackage{orcidlink}
\usepackage{newtxtext,newtxmath}
\usepackage{booktabs}
\usepackage{upgreek}
\usepackage{ulem}
\usepackage{threeparttable}
\usepackage[T1]{fontenc}
\usepackage{comment}
\DeclareRobustCommand{\VAN}[3]{#2}
\let\VANthebibliography\thebibliography
\def\thebibliography{\DeclareRobustCommand{\VAN}[3]{##3}\VANthebibliography}
\usepackage{graphicx}	
\usepackage{amsmath}	
\usepackage{tablefootnote}
\usepackage{url}
\bibpunct{(}{)}{;}{a}{}{,} 
\usepackage{txfonts}
\usepackage{lscape}
\usepackage{rotating}
\usepackage{hyperref}
\hypersetup{colorlinks=true,linkcolor=blue,citecolor=blue,filecolor=blue}

\begin{document}

\title{Radio ghosts: Radio polarimetry study of newly discovered cluster minihalo candidates or radio tailed galaxies}
\titlerunning{Radio polarimetry study of cluster minihalo candidates or radio tailed galaxies}

   \author{Sam Taziaux \inst{\ref{rub},\ref{rapp}, \ref{csiro}}\orcidlink{0009-0001-6908-2433}
   \and 
   Christopher J. Riseley\inst{ \ref{rub},\ref{rapp}}\orcidlink{0000-0002-3369-1085}
   \and
    Stefan W. Duchesne \inst{\ref{csiro}}\orcidlink{0000-0002-3846-0315}
    \and
    Francesca Loi \inst{\ref{inaf}}\orcidlink{0000-0002-8627-6627}
    \and
   Alec J. M. Thomson \inst{\ref{csiro},\ref{ska}}\orcidlink{0000-0001-9472-041X}
   \and    
   Aritra Basu \inst{\ref{dza},\ref{mpifr},\ref{tls}}\orcidlink{0000-0003-2030-3394}
   \and             
   Timothy J. Galvin \inst{\ref{csiro}}\orcidlink{0000-0002-2801-766X}
   \and 
   Dominik J. Bomans \inst{\ref{rub}, \ref{rapp}}\orcidlink{0000-0001-5126-5365}
    }
    
   \institute{Ruhr University Bochum, Faculty of Physics and Astronomy, Astronomical Institute (AIRUB), Universitätsstraße 150, 44801 Bochum, Germany
              \email{sam.taziaux@rub.de} \label{rub}
        \and 
             Ruhr Astroparticle and Plasma Physics Center (RAPP Center) \label{rapp}
        \and 
            CSIRO Space and Astronomy, PO Box 1130, Bentley WA 6102, Australia \label{csiro}
        \and 
            INAF – Osservatorio Astronomico di Cagliari, Via della Scienza 5, Selargius, Italy \label{inaf}
        \and 
            SKA Observatory, SKA-Low Science Operations Centre, 26 Dick Perry Avenue, Kensington WA 6151, Australia \label{ska}
        \and
            Deutsches Zentrum f\"ur Astrophysik, Postplatz 1, 02826 G\"orlitz, Germany \label{dza}
        \and
            Max Planck Institut f\"ur Radioastronomie, Auf den H\"ugel 69, 53121 Bonn, Germany \label{mpifr}
        \and
            Th\"uringer Landessternwarte, Sternwarte 5, 07778 Tautenburg, Germany \label{tls}}

\date{Accepted XXX. Received YYY; in original form ZZZ}



\abstract
{Extended diffuse radio emission in the central regions of galaxy clusters is commonly associated with non-thermal processes in the intracluster medium (ICM), but its origin and the mechanisms responsible for sustaining it remain uncertain.}
{We present the discovery and multi-wavelength analysis of two diffuse radio sources, MKT\,J0355--4237 and MKT\,J0355--4230, to assess their nature and investigate their radio spectral and magnetic field properties through radio polarimetric analysis.}
{We use deep MeerKAT L-band (1.28\,GHz) observations to perform spectral analysis of the diffuse emission and study its polarisation properties via rotation measure synthesis. These are complemented by eROSITA X-ray data and DESI photometric redshift catalogues to characterise the host cluster environments.}
{MKT\,J0355--4237 is confirmed as cluster-scale minihalo with extent of $\sim 400$\,kpc, while the other diffuse source MKT\,J0355--4230 remain uncertain and could be either associated with a radio minihalo or radio tailed galaxy. Both show steep spectra ($\alpha \approx -0.7$ to $-0.8$) that steepen toward the outskirts ($\alpha \lesssim -2.5$), consistent with spatial variations in electron ageing and re-acceleration. Polarised emission is detected in both systems, with fractional polarisation $\sim1\,\%$ in the centres and increasing in the outer regions of MKT\,J0355--4237. MKT\,J0355--4230 is associated with a previously uncatalogued cluster and shows an X-ray morphology offset from the brightest cluster galaxy, indicative of dynamical activity. MKT\,J0355--4237 exhibits a sharply bounded internal radio deficit without a corresponding X-ray counterpart, within current sensitivity limits.}
{The spectral steepening supports turbulent re-acceleration as the origin of the diffuse emission. The detection of polarisation indicates partially ordered magnetic fields in the ICM. The internal radio deficit may trace local variations in particle re-acceleration or fossil AGN plasma embedded within the minihalo.}


   \maketitle
%

\section{Introduction}
Galaxy clusters hosting cool cores\footnote{Systems with dense, X-ray bright centres, which are characterised by short radiative cooling times and declining central temperature profiles \citep[e.g.][]{Hudson_2010}.} frequently exhibit diffuse synchrotron radio emission, termed radio `minihalos', which typically envelop the brightest cluster galaxy (BCG) and permeate the innermost tens to a few hundred kiloparsecs of the intracluster medium (ICM), without any direct morphological connection to active galactic nucleus (AGN) jets or lobes \citep{Govoni_2004, Giacintucci_2011, Gitti_2016}.
These sources are characterised by steep radio spectra ($S_\nu \propto \nu^{\alpha_{\rm nth}}$ with $\alpha_{\rm nth} < -1$), low surface brightness, and relatively extended morphologies \citep[$\sim$50–300\,kpc in most systems, reaching up to $\sim$500 kpc in extreme cases;][]{Govoni_2004, Giacintucci_2014, Doria_2012, Bonafede_2023, Riseley_2023}.
Linear polarisation has generally remained undetected in radio minihalos, with only a few reported cases, such as the detection in the minihalo of RBS\,797 \citep{Gitti_2006}. The lack of widespread polarisation detections is commonly attributed to the presence of tangled magnetic fields and strong internal Faraday depolarisation within the dense cool-core regions \citep{Govoni_2004}.

Two main framework has been proposed to explain the origin of the radio emission in minihalos, namely turbulent re-acceleration scenario \citep[e.g.][]{Gitti_2002} and hadronic model \citep[e.g.][]{Pfrommer_2004}.
In the turbulent re-acceleration model, subsonic gas motions in cool-core clusters, generated by the sloshing of the dense central gas within the cluster's gravitational potential, induce turbulence that both re-accelerates relativistic electrons and amplifies the magnetic field. These `sloshing' motions are typically triggered by minor mergers or gravitational perturbations from infalling subclusters or galaxies, which displace the dense core gas from the cluster center and induce oscillatory motions within the cluster potential \citep{Giacintucci_2014}.
This sloshing process naturally produces spiral-shaped X-ray cold fronts, which are sharp boundaries between cool dense gas and the surrounding hotter ICM.
The close morphological correspondence between the radio emission and X-ray cold fronts observed in several systems \citep[e.g. RXC\,J1504.1--0248;][]{Giacintucci_2011, Giacintucci_2024} provides strong empirical support for this scenario.
This connection is further supported by the observed correlation between minihalo radio power and the X-ray luminosity of the host cluster, suggesting a close link between the thermal properties of the cool-core ICM and its non-thermal emission \citep{Ignesti_2020}.
In contrast, the hadronic (or secondary) model attributes the relativistic electrons to inelastic collisions between long-lived cosmic-ray protons and the thermal protons of the ICM, which continuously inject secondary electrons and positrons that radiate synchrotron emission \citep{Pfrommer_2004}.
Although this model naturally provides a persistent population of relativistic particles, it is constrained by $\gamma$-ray upper limits and the required cosmic-ray proton energy budget in some systems \citep{Doria_2012}. Nevertheless, simulations show that secondary electrons produced in sloshing cool-core environments can reproduce minihalo-like radio emission \citep{ZuHone_2015}.

Magnetic field strengths in galaxy clusters are commonly inferred from Faraday rotation measure (RM) analysis of polarised radio sources seen through the ICM \citep[e.g.][]{Carilli_2002, Govoni_2004, Feretti_2012, Reiss_2014}. These studies indicate that non-cool-core clusters generally host comparatively weak and less structures magnetic fields ($\sim1-10\,\upmu$G), whereas cool-core cluster systems often require central field strengths of $10-40\,\upmu$G to reproduce the observed RM dispersions, with the field strength declining steeply with radius \citep{Govoni_2004, Feretti_2012, Ignesti_2020}. The magnetic fields in cool-core centres are therefore inferred to be relatively strong and ordered on scales of several kiloparsecs.

Recent observational advances have extended our understanding of minihalos to earlier cosmic epochs.
The discovery of a candidate radio minihalo in the massive cool-core galaxy cluster SpARCS\,104922.6+564032.5 at $z=1.709$ with LOFAR \citep{HLavacek-Larrondo_2025} demonstrates that diffuse synchrotron emission and dynamically significant magnetic fields were already in place when the Universe was younger than 4\,Gyr.
The measured radio power, morphological alignment with the X-ray emitting ICM, and inferred central magnetic field strengths ($\gtrsim10\,\upmu$G) suggest that synchrotron losses may remain competitive with inverse-Compton losses against the cosmic microwave background (CMB), despite the stronger CMB energy density expected at high redshift \footnote{Since the CMB energy density scales as $U_{\rm CMB} \propto (1+z)^4$, inverse-Compton losses become increasingly important at high redshift, although strong magnetic fields can keep synchrotron losses competitive \citep[e.g.][]{Murphy_2009}.}
Large-sample statistical studies \citep[e.g.][]{Giacintucci_2014, Giacintucci_2017} have established that nearly all massive, hot, and relaxed clusters with cool cores host minihalos, and that these are almost exclusively confined within regions affected by gas sloshing, reinforcing the physical link between cool-core dynamics, turbulence, and non-thermal phenomena \citep{Biava_2024}.

Understanding radio minihalos in cool-core clusters is essential for constraining the coupling between thermal and non-thermal components in cluster cores. In these systems, minihalos are uniquely associated with dense, rapidly cooling ICM and are thought to arise from the re-acceleration of relativistic particles by turbulence driven by core sloshing and/or AGN feedback. They therefore provide a direct probe of how turbulence, magnetic fields, and cosmic-ray populations evolve in high-pressure environments where radiative cooling and feedback are closely linked.
At higher redshift, where cool-core formation and AGN feedback operate under different conditions, minihalos can further constrain the evolution of intracluster magnetic fields and the efficiency of particle acceleration in regimes where inverse-Compton losses are enhanced. Progress in this field requires sensitive, high-resolution radio observations of cool-core clusters, together with spatially resolved X-ray data to characterise sloshing signatures and the thermal structure of the core.

In this study, we focus on two newly discovered diffuse radio sources and investigate their observed properties to assess whether they are consistent with minihalo emission.
The paper is structured as follows. In Sect.~\ref{data}, we describe the observational data and the procedures used for its reduction. We show the identification of the associated galaxy clusters in Sect.~\ref{cluster_ident}. In Sect.~\ref{results}, we present the results of the radio polarimetric analysis, along with complementary X-ray findings. Sect.~\ref{discussion} discuss the radio feature in the minihalo and disentangle turbulent from ordered magnetic field in the ICM. Finally, Sect.~\ref{conclusion} summarises our main conclusions.

\section{MeerKAT observations and data}
\label{data}
\begin{figure*}
        \centering
    \includegraphics[width=0.98\linewidth]{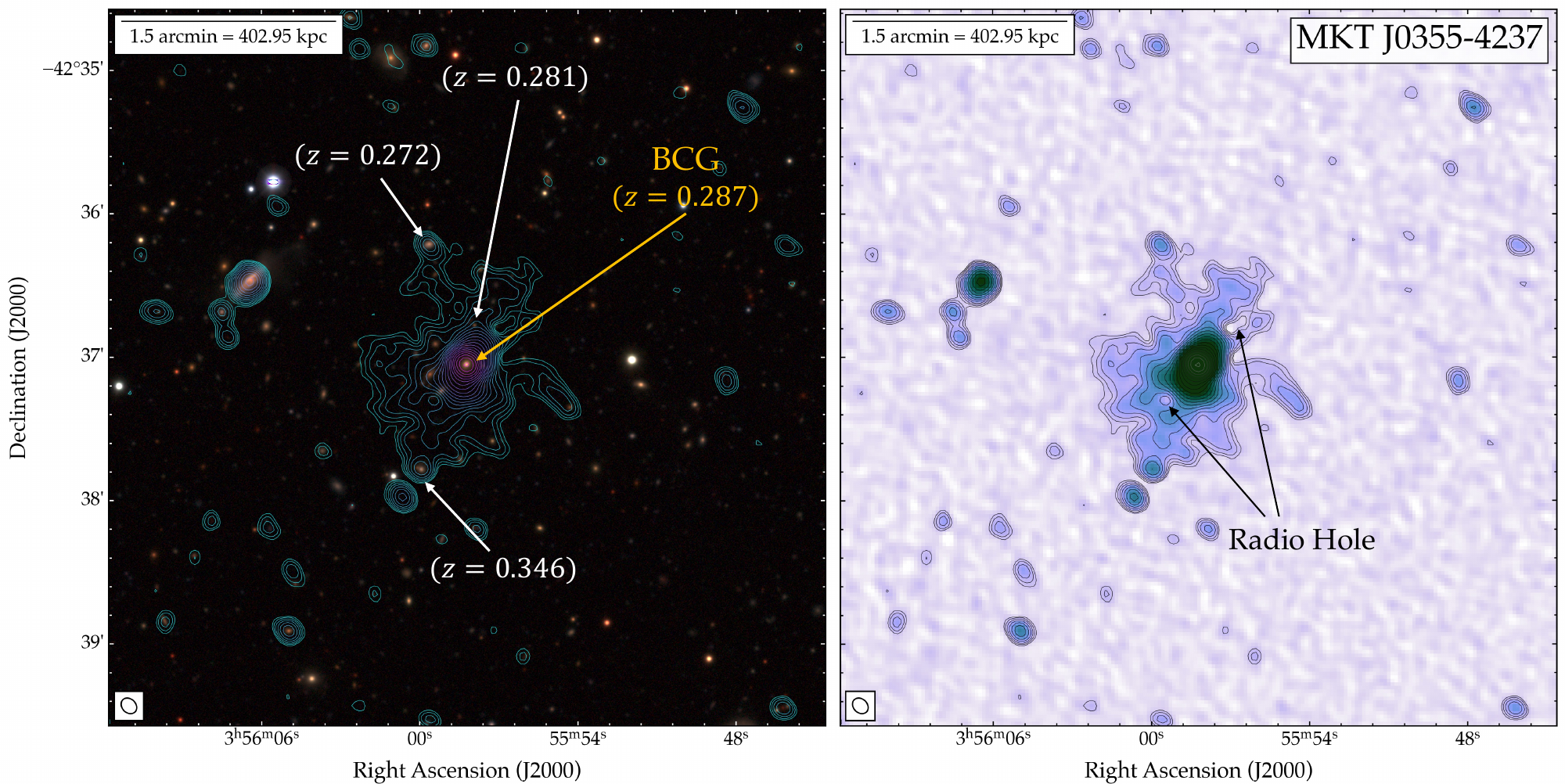}
    \caption{Colour-composite image from the Legacy Imaging Surveys (z, r and g bands) of MKT\,30355--4237 with overlaid MeerKAT L-band radio emission contours starting at $3\,\sigma$ and increasing by a factor of $\sqrt{2}$ at a central frequency of 1.28\,GHz with a noise level $\sigma=6.5\,\upmu$Jy/beam. 
    \textit{Right:} MeerKAT L-band radio emission with overlaid contours starting at $3\,\sigma$ and increasing by a factor of $\sqrt{2}$ at a central frequency of 1.28\,GHz with a noise level $\sigma=6.5\,\upmu$Jy/beam. The scale on the top left have been calculate using the redshift of the BCG in the center of the minihalo candidate.
    The beam of $7.23\arcsec \times 5.75\arcsec$ is shown in the bottom left corner.
    }
    \label{desi_ghost}
\end{figure*}

\begin{figure*}
        \centering
    \includegraphics[width=0.98\linewidth]{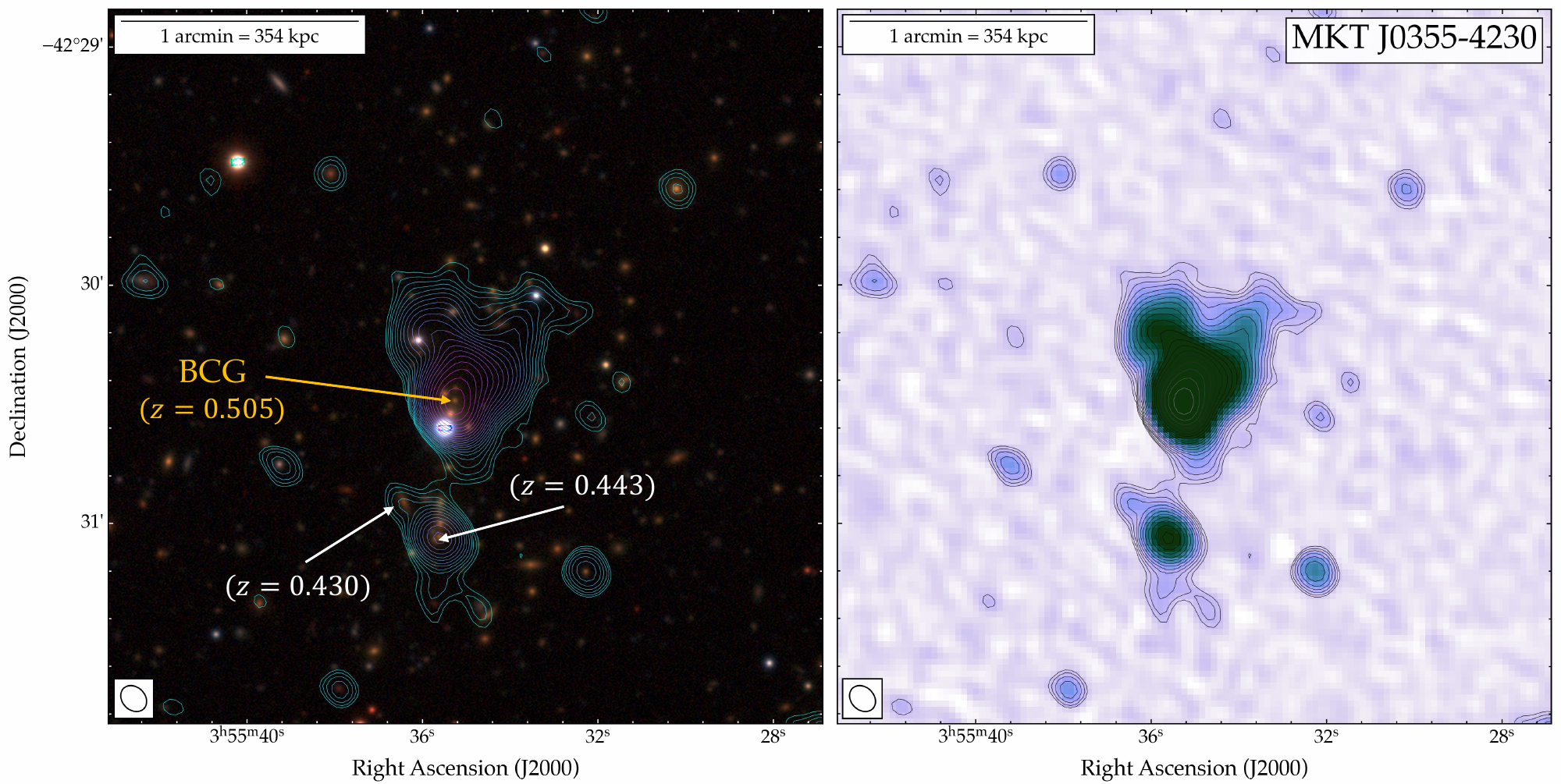}
    \caption{Colour-composite image from the Legacy Imaging Surveys (z, r and g bands) of MKT\,J0355--4230 with overlaid MeerKAT L-band radio emission contours starting at $3\,\sigma$ and increasing by a factor of $\sqrt{2}$ at a central frequency of 1.28\,GHz with a noise level $\sigma=6.5\,\upmu$Jy/beam. 
    \textit{Right:} MeerKAT L-band radio emission with overlaid contours starting at $3\,\sigma$ and increasing by a factor of $\sqrt{2}$ at a central frequency of 1.28\,GHz with a noise level $\sigma=6.5\,\upmu$Jy/beam. The scale on the top left have been calculate using the redshift of the BCG in the center of the minihalo candidate.
    The beam of $7.23\arcsec \times 5.75\arcsec$ is shown in the bottom left corner.
    }
    \label{desi_heart}
\end{figure*}

\subsection{Observation}
The two minihalo candidates MKT\,J0355--4237 (Fig.~\ref{desi_ghost}) and MKT\,J0355--4230 (Fig.~\ref{desi_heart}) were observed by MeerKAT on 12-October-2023 and 15-October-2023 under project ID SCI-20230907-ST-01, PI: S. Taziaux. The observations were conducted in L-band (1.28\,GHz, 856\,MHz bandwidth) with a total integration time of 6\,hrs. Although these sources were not the primary target of these observations, they lie at an angular distance of $14.98\arcmin$ and $8.34\arcmin$ from the pointing centre of the observed field for each observation, which is still (1) within the primary beam FWHM of MeerKAT L-band and (2) within the radius where MeerKAT's off-axis polarisation is negligible \citep[][]{Sekhar_2022}.
This corresponds to a polarisation leakage of only $\sim 0.02\%$ at the position of the most distant minihalo candidate for frequencies above 1.4\,GHz \citep{HugoPerley_2024}

\subsection{Data reduction}
\label{datameerkat}
The MeerKAT observations at L-band, covering the frequency range 856 to 1712\,MHz, were recorded using 32768 frequency channels. These data were
averaged to 4096 209.984\,kHz-wide channels for calibration at a central 
frequency of 1284\,MHz. On both days of observations, 
J$0408\textrm{--}6545$ was used to set the
absolute flux density scale, solve the bandpass, and because it is weakly polarised, it was also used to solve the on-axis instrumental leakage. The complex gains were solved
using an unresolved bright source, J$0440\textrm{--}4333$, and the absolute polarisation
was calibrated using 3C\,138. J$0440\textrm{--}4333$ show a calibration leakage of $0.27\,\%$. Calibration were performed following standard
procedure using the \texttt{CASA package} \citep[][]{casa}, wherein, the absolute flux scale for J$0408\textrm{--}6545$ was set using the task \texttt{setjy} following the strategy provided by the SARAO,\footnote{\url{https://skaafrica.atlassian.net/wiki/spaces/ESDKB/pages/1481408634/Flux+and+bandpass+calibration}} and the polarisation
model for 3C\,138 was adopted from \cite{PerleyButler2013}. We adopt a 5\,\% uncertainty on the absolute flux-density scale.

Iterative imaging and self-calibration were performed using \textsc{WSClean 3.0} \citep{wsclean} and \texttt{CASA 6.4.4.31} (phase-only, frequency-independent calibration with solution intervals ranging from 2.5\,min down to 10\,s) until convergence, defined as no significant improvement in image dynamic range and image RMS between successive iterations, together with stable gain solutions. 
Imaging was performed using Briggs weighting with $\texttt{robust} = -1, -0.5, -0.3, 0, 0.3$ to slowly reconstruct the diffuse emission. 
Self-calibration was performed independently for the two days of observations, and by splitting the 856-MHz bandwidth into 32 sub-bands of 26.75\,MHz each, to avoid bandwidth smearing larger than the synthesised beam, during imaging of the wide field of view.
Multifrequency and multiscaling CLEANing \citep{hoegbom_1974,Offringa_2017_MS} utilising interactive masks using \texttt{flint\_masking}~(Galvin et al. in prep.)\footnote{\url{https://github.com/flint-crew/flint}} around visible sources to minimise artefacts and applying detection thresholds to retain genuine emission only. The deconvolution was performed down to a threshold of $0.5\sigma$.
After the final round of self-calibration, final imaging in all Stokes parameters were done by combining the two datasets in  \textsc{WSClean}.
The total intensity continuum image was generated using Briggs robust weighting with $\texttt{robust} = -2$, with \texttt{-no-mf-weighting} in the
last round of imaging to ensure not to include artificial spectral
substructures, resulting in resolutions of $7.23\arcsec \times 5.75\arcsec$, and rms of $6.5\,\upmu$Jy/beam limited by the confusion noise. 
The primary beam have been corrected using \texttt{-apply-primary-beam} task in \textsc{WSClean}.

For polarisation imaging, we individually imaged the Stokes~$Q$ and $U$ parameters with a channel width of 13.5\,MHz to mitigate bandwidth depolarisation effects at a Briggs robust weighting of $\texttt{robust}=-1$. We used a higher positive robust weighting for the Stokes~$Q$ and Stokes~$U$ imaging than for Stokes~$I$ in order to enhance the sensitivity to diffuse polarised emission.  
In order to optimise per channel deconvolution of Stokes~$Q$ and Stokes~$U$, we used \texttt{-squared-channel-joining} and \texttt{-join-polarizations} parameters in \textsc{WSClean}.
The resulting 39 images have been primary beam corrected and convolved to lowest common resolution of $9\arcsec$ to perform RM synthesis \citep[][]{Burn_1966, Brentjens_2005, Heald_2009_II} using the \textsc{RMtools}\footnote{\url{https://github.com/CIRADA-Tools/RM-Tools}} \citep[][]{rmtools} suite task \texttt{rmsynth3d}, sampling between $-698.79$ and $698.79\,\,\mathrm{rad}\,\mathrm{m}^{-2}$. The Faraday depth axis was oversampled by a factor of $20$ relative to the RMSF FWHM, resulting in $469$ sampled Faraday depth channels.
To deconvolve the rotation measure spread function (RMSF) from the Faraday dispersion function (FDF), thereby reducing sidelobes and mitigating artificial structure, we perfrom RM-clean using the \textsc{RMtools} suite task \texttt{rmclean3d}. 
The polarised intensity ($PI$) was derived as the maximum value along the Faraday axis of the polarised intensity cube, calculated from $PI = \sqrt{Q^2 + U^2}$ \citep[][]{Wardle_1974}.
The RMSF has a FWHM of $59.72\,\mathrm{rad\,m^{-2}}$, with sensitivity to Faraday structures up to scales of $\sim 100\,\mathrm{rad\,m^{-2}}$. The sampled Faraday depth range of $\pm 698.79\,\mathrm{rad\,m^{-2}}$ is well within the instrumental maximum observable range and does not limit the analysis\footnote{The maximum recoverable Faraday scale is $\delta \phi_{\max} \approx \pi / \lambda_{\min}^2 \approx 103\,\mathrm{rad\,m^{-2}}$, where $\lambda_{\min}$ corresponds to the highest observed frequency (1712\,MHz). The maximum observable Faraday depth is set by the channel width in $\lambda^2$ space and is $\gg 10^4\,\mathrm{rad\,m^{-2}}$, and therefore not limiting for this analysis.}.
When performing RM-clean, we cleaned down to a threshold of $5\sigma_{QU}$ ($\sigma_{QU}=2.0\,\upmu$Jy/beam). The noise levels were estimated from a slice through the Stokes $Q$ and $U$ maps at a Faraday depth, and the uncertainty $\sigma_{QU}$ was then derived through standard error propagation as $\sigma_{QU} = \sqrt{\frac{Q^2}{Q^2+U^2}\sigma_Q^2+\frac{U^2}{Q^2+U^2}\sigma_U^2}$. We perform the \texttt{rmtools\_peakfitcube} task, taking a 3D Faraday depth cube, analysing the RM spectrum at each spatial pixel, fitting the main peaks in the FDF, and correcting for the Ricean-bias\footnote{Peak polarised intensity after Ricean-bias correction, applied when snrPIfit > 5, using $PI_{\rm eﬀ} = \sqrt{PI^2-2.3\,\sigma^2_{\rm FDF}}$ \citep{George_2012}.}.
Furthermore, the background polarised intensity level was estimated from emission-free Faraday depth channels away from the fitted peak and subtracted from the measured polarised intensity.

The polarisation angle $\chi$ was determined from the Stokes parameters using $\chi = \frac{1}{2} \arctan\left(\frac{U}{Q}\right)$, providing the orientation of the electric field vectors in the plane of the sky. Under the assumption that the observed RM and polarised emission originate from the same region, the vectors were derotated to recover the intrinsic magnetic field orientation via $\chi_0 = \chi - \text{RM} \, \lambda_0^2 + \frac{\pi}{2}$, where $\lambda_0^2$ is the weighted mean of the observed $\lambda^2$ values using the channel weights employed in RM synthesis derived from the Stokes~$Q$ and $U$ noise. For these data, we find $\lambda_0^2 = 0.059\,\mathrm{m}^2$, corresponding to $1226$\,MHz, which was used as the reference wavelength in the derotation.
The RM describes the observed rotation measure corrected
for Faraday rotation by the Galactic foreground. We used the most recent version of the Galactic RM map \citet{Hutschenreuter_2023} to estimate the Galactic contribution, being RM$_{\rm Gal}=3.54\pm0.92$\,rad\,m$^{-2}$. The maps of $\chi_0$ presented herein have also been corrected for this Galactic Faraday rotation.

\section{General Properties}
\label{cluster_ident}
These MeerKAT data reveal two new discovered diffuse source with a bright core and diffuse emission reaching into the halo with multiple galaxies, which we call a minihalo candidate. The first minihalo candidate MKT\,J0355--4237, shown in Fig.~\ref{desi_ghost} is located at RA $= 03^\text{h}\,55^\text{m}\,58.492^\text{s}$ and Dec $= -42^\text{d}\,37^\text{m}\,08.967^\text{s}$. The second minihalo candidate MKT\,J0355--4230, shown in Fig.~\ref{desi_heart} is located at RA $= 03^\text{h}\,55^\text{m}\,35.215^\text{s}$ and Dec $= -42^\text{d}\,37^\text{m}\,29.665^\text{s}$. 

\subsection{Identification of a galaxy cluster}
Galaxy clusters provide the environment required for the formation of radio minihalos, as these sources are typically associated with the dense ICM and the central regions of relaxed systems.
To assess whether the candidate minihalos are associated with galaxy clusters, we first crossmatched their positions with existing galaxy cluster catalogues \citep{Wen_2024}. MKT\,J0355--4237 is spatially coincident with the catalogue cluster CFSFDP\,J035558.2--423703 at a redshift of $z_{\rm cl}=0.2854$. The cluster has a reported richness of $\lambda=12.1$, corresponding to nine candidate member galaxies within $r_{500}=0.513$\,Mpc and an estimated mass of $M_{500}=5.1\times10^{13}\,M_\odot$, indicating a relatively poor cluster or massive group environment. The photometric redshift of the BCG, $z_{\rm photo}=0.287\pm0.020$, is in excellent agreement with the catalogue redshift, confirming that the diffuse radio emission is associated with a previously identified galaxy cluster. The BCG has a stellar mass of $\log(M_\star/M_\odot)=11.28$.

\begin{figure*}
    \centering
    \includegraphics[width=0.42\linewidth]{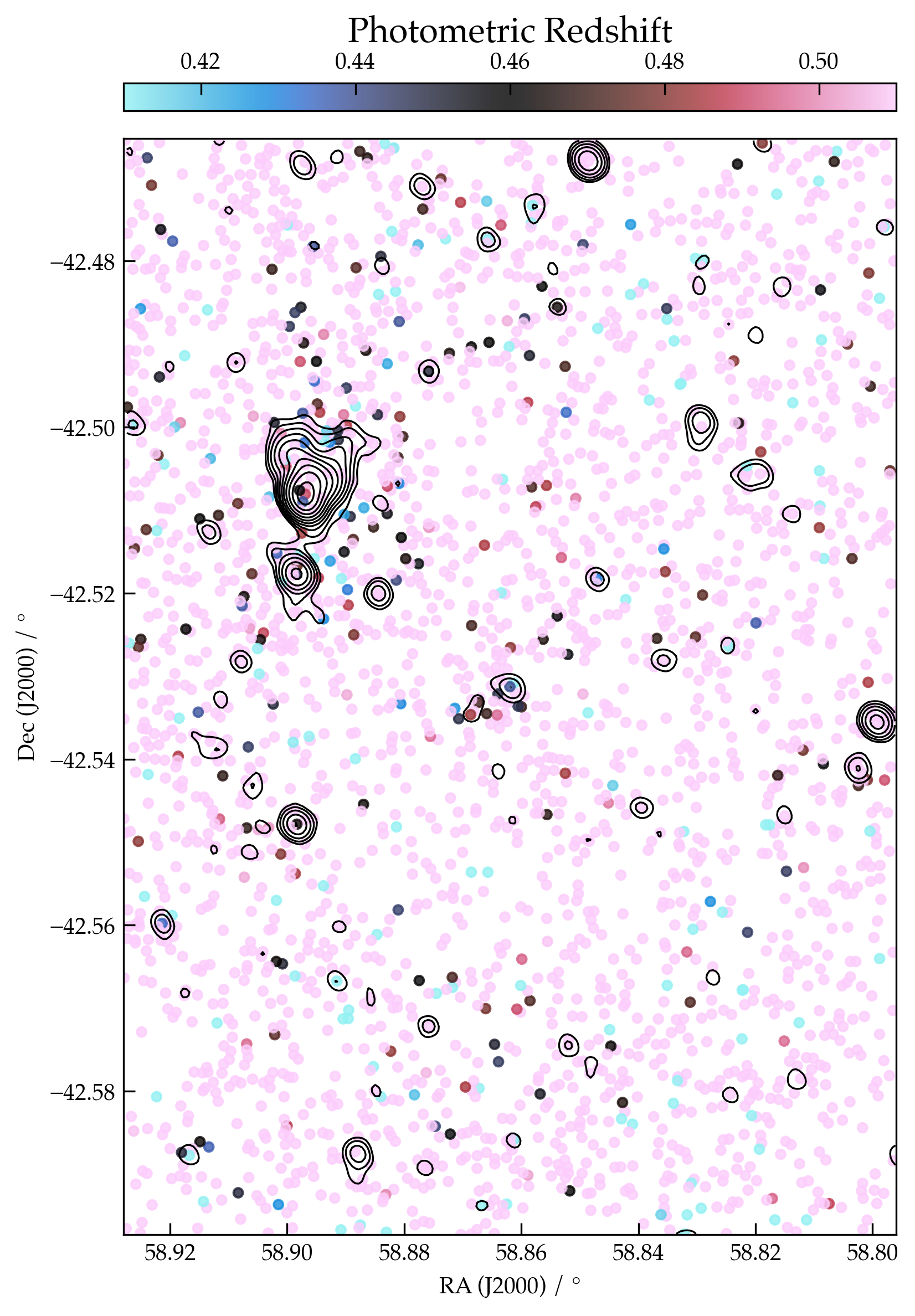}
    \includegraphics[width=0.425\linewidth]{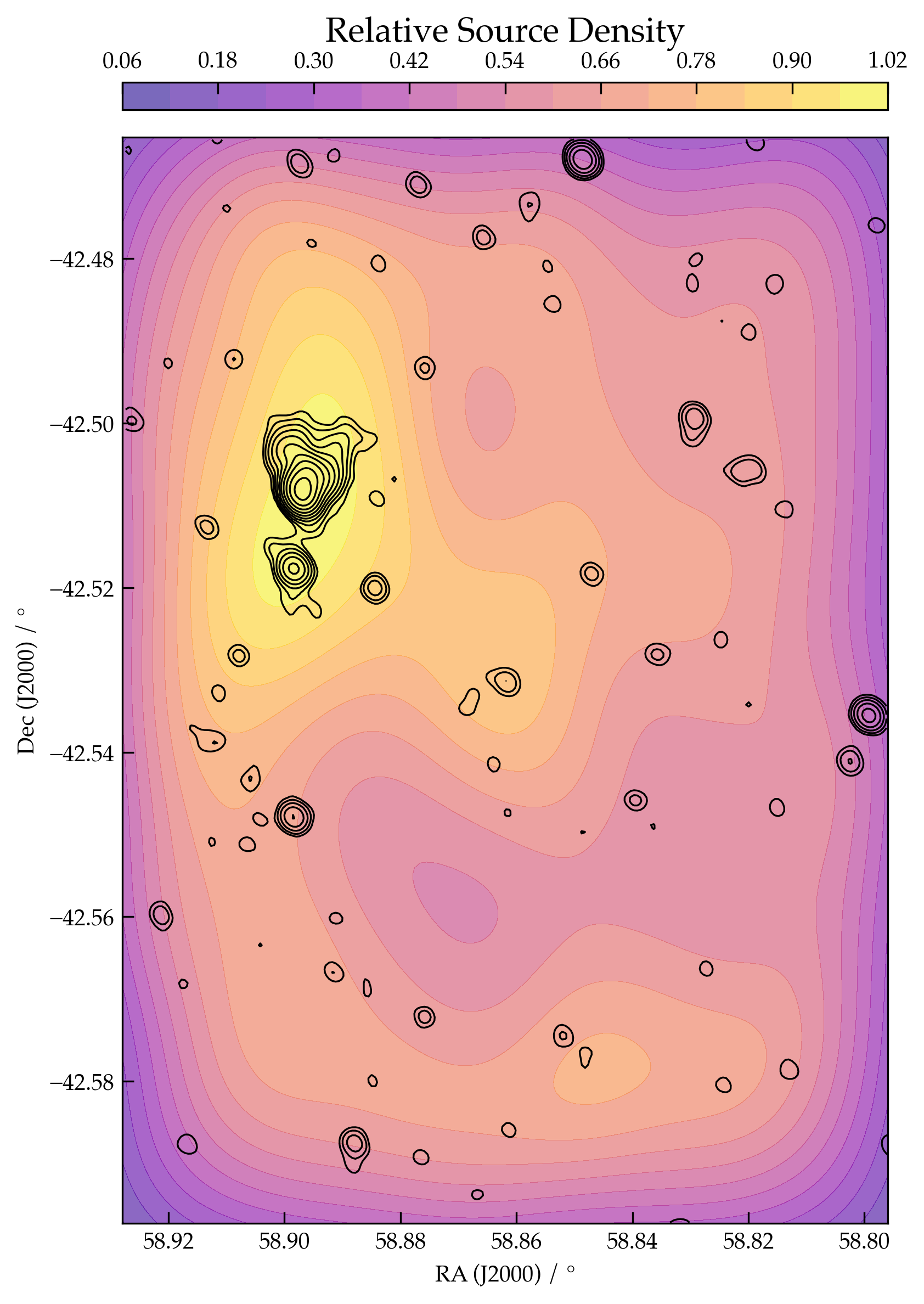}    
    \caption{Galaxy distribution based on photometric redshifts. \textit{Left panel:} Photometric redshift distribution of galaxies in the field of MKT\,J0355--4230. \textit{Right panel:} Kernel density estimate of the projected galaxy distribution, illustrating the relative source density in the field. The distributions are derived from DESI photometric redshift catalogues \citep{Zhou_2023}. All the panels are overlaid with MeerKAT L-band total intensity contours starting at $3\sigma$ and increasing by a factor of 2 ($\sigma=6.5\,\upmu$Jy/beam).}
    \label{minihalo_envir}
\end{figure*}
For MKT\,J0355--4230, no clear cluster counterpart is found at the position of the diffuse radio emission itself. However, the source lies in the vicinity of a redMaPPer cluster candidate centred approximately $2\arcmin$ to the north-west of the radio source. At the cluster redshift of $z_{\rm cl}=0.4632$, this angular separation corresponds to a projected distance of $\sim720$\,kpc. The redMaPPer system has a characteristic radius of $R_{\lambda}=912$\,kpc and a richness of $\lambda=72.7$, implying that the radio source is located within the expected extent of the cluster environment.
The candidate BCG associated with the diffuse emission has a photometric redshift of $z_{\rm photo}=0.505\pm0.070$. The difference between the nominal BCG and cluster redshifts is $\Delta z = 0.0418$, which corresponds to a comoving-distance offset of approximately $195$\,Mpc. At face value this would suggest that the BCG lies well behind the redMaPPer cluster. However, the uncertainty on the photometric redshift is large ($\sigma_z=0.070$), corresponding to a comoving-distance uncertainty of approximately $230$\,Mpc, exceeding the inferred separation between the two systems. The estimated mass of the cluster is $M_{500}=3.2\times10^{14}\,M_\odot$.

To further investigate whether MKT\,J0355--4230 is associated within the redMaPPer system, we analyse the galaxy density distribution in the surrounding fields using photometric redshifts from the DESI catalogue \citep{Zhou_2023}. 
In Fig.~\ref{minihalo_envir}, we show the cluster BCG at redshift $z_{\rm cl}=0.4632$ at the center and showing the 912\,kpc ($\approx 2.5\arcmin$). In the left panel of Fig.~\ref{minihalo_envir}, we show the redshift distribution and the right panel of Fig.~\ref{minihalo_envir} show the kernel density estimates within the redMaPPer galaxy cluster. We show a confidence interval of $0.42 \le z_{\rm cl} \le 0.50$.
The clear overdensity of galaxies seen in the DESI photometric-redshift distribution and the fact that the diffuse radio emission lies within the projected extent of the redMaPPer system, the available evidence favours an association between MKT\,J0355--4230 and this galaxy cluster. Nevertheless, spectroscopic redshifts will be required to firmly establish the cluster membership and confirm the nature of the diffuse radio source.

\section{Properties of the radio polarimetry study}
\label{results}



\begin{figure*}
    \centering
    \includegraphics[width=0.95\linewidth]{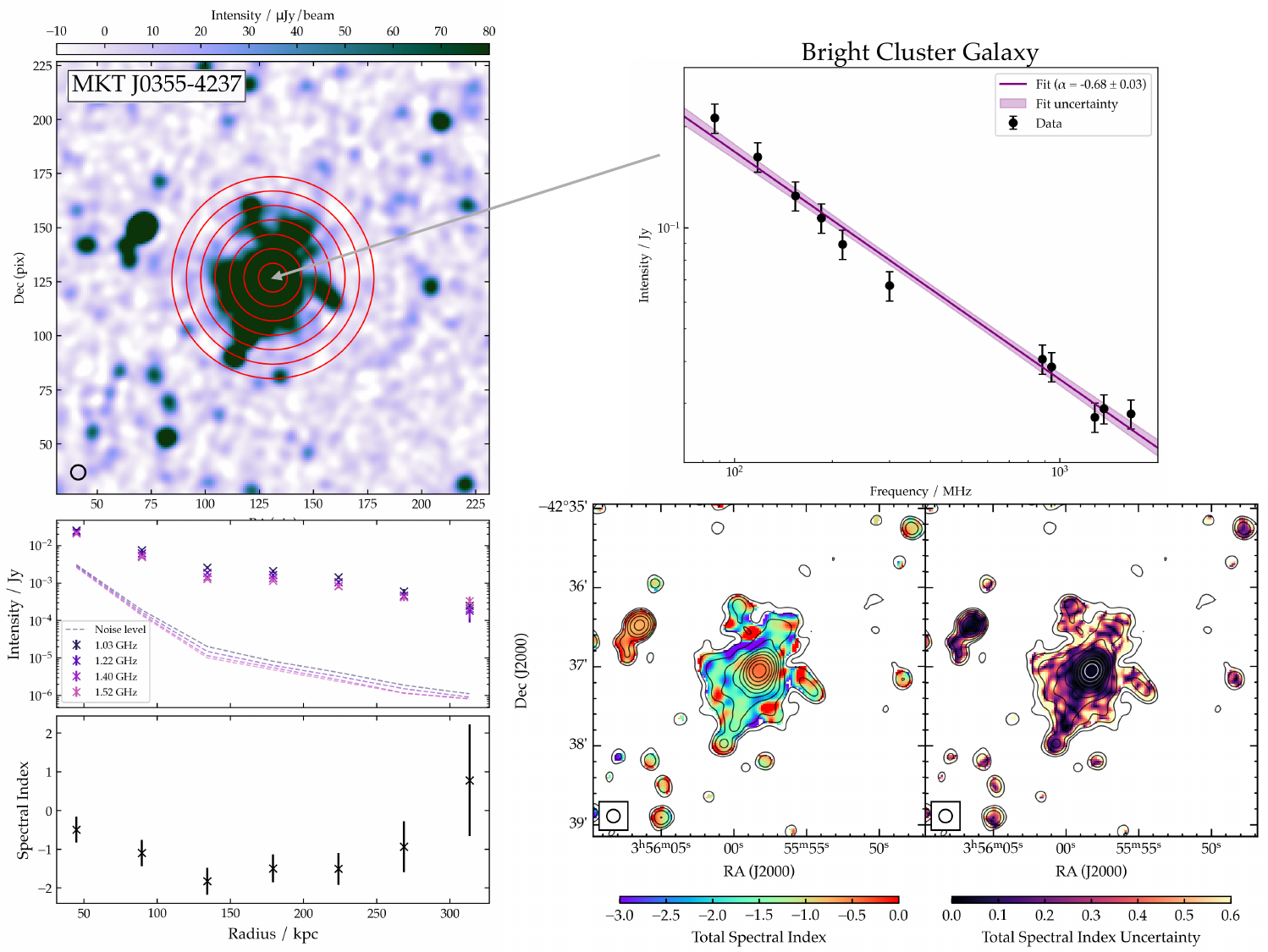}
    \caption{The MeerKAT L-band total intensity map at 1.28\,GHz is shown with the aperture in red indicating the individual bins with the the total intensity profiles at 1.03, 1.22, 1.40 and 1.52\,GHz are presented together with the corresponding spectral index profile. The dashed lines indicate the noise level measured in each annulus at the corresponding frequency (left panel). Integrated spectral energy distribution for the unresolved radio BCG using the total intensity data with different observations: GLEAM-X data \citep{gleamx}, GLEAM-300 data \citep{Duchesne_2025_gleam300}, RACS-Low data \citep{racs}, ASKAP EMU main survey \citep{emu_main}, MeerKAT 1.28\,GHz data (this work), RACS-Mid \citep{racsmid}, and RACS-High \citep{racshigh}. A simple power-law model have been fitted to the data points (right top panel). An inband total spectral index of the MeerKAT L-band and its uncertainty between the frequencies 1.03 and 1.52 GHz overlaid with radio emission at frequency of 1.28 GHz starting at $3\sigma$ and increasing by a factor of 2 ($\sigma = 8 \upmu$Jy/beam). The resolution of $10\arcsec$ is shown in the bottom left corner (right bottom panel).}
    \label{fig_minihalo_ghost_spectra}
\end{figure*}

\begin{figure*}
    \centering
    \includegraphics[width=0.95\linewidth]{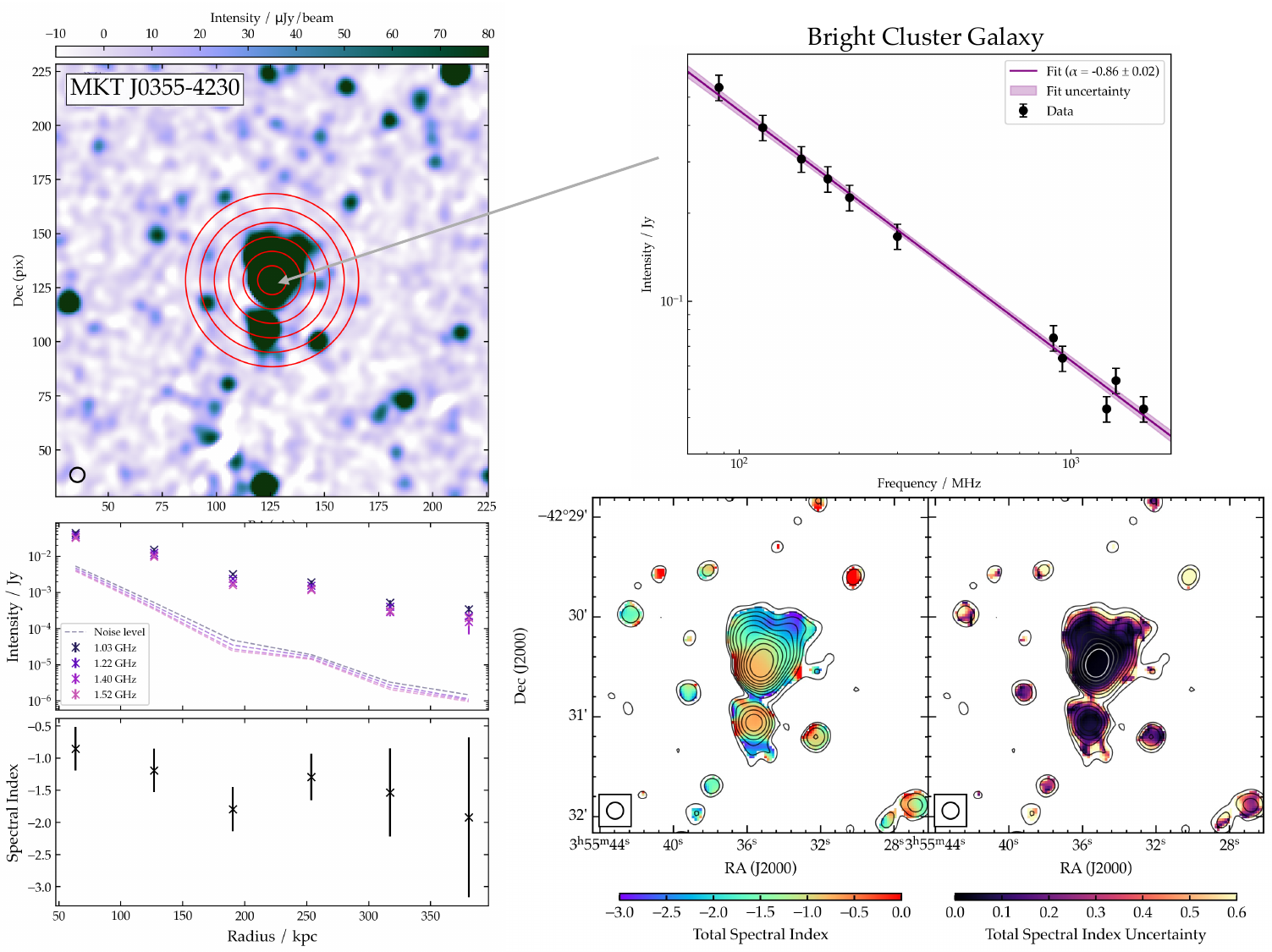}
    \caption{The MeerKAT L-band total intensity map at 1.28\,GHz is shown with the aperture in red indicating the individual bins with the the total intensity profiles at 1.03, 1.22, 1.40 and 1.52\,GHz are presented together with the corresponding spectral index profile. The dashed lines indicate the noise level measured in each annulus at the corresponding frequency (left panel). Integrated spectral energy distribution for the unresolved radio BCG using the total intensity data with different observations: GLEAM-X data \citep{gleamx}, GLEAM-300 data \citep{Duchesne_2025_gleam300}, RACS-Low data \citep{racs}, ASKAP EMU main survey \citep{emu_main}, MeerKAT 1.28\,GHz data (this work), RACS-Mid \citep{racsmid}, and RACS-High \citep{racshigh}. A simple power-law model have been fitted to the data points (right top panel). An inband total spectral index of the MeerKAT L-band and its uncertainty between the frequencies 1.03 and 1.52 GHz overlaid with radio emission at frequency of 1.28 GHz starting at $3\sigma$ and increasing by a factor of 2 ($\sigma = 8 \upmu$Jy/beam). The resolution of $10\arcsec$ is shown in the bottom left corner (right bottom panel).}
    \label{fig_minihalo_heart_spectra}
\end{figure*}

\subsection{Radio continuum properties}
In the right panels of Figs.~\ref{desi_ghost} and \ref{desi_heart}, we present the MeerKAT 1.28\,GHz radio continuum emission of two diffuse sources, which we consider as candidate minihalos and discuss in the following sections.
In Fig.~\ref{desi_ghost}, a bright central source is detected at the peak of the diffuse emission, which we identify as the BCG, associated with WISEA\,J035558.23--423703.0 at a redshift of $z_{\rm photo} = 0.287\pm0.020$. 
Assuming that the diffuse radio emission is associated with this source, the physical extent of the candidate minihalo is approximately $400$\,pc. Several compact sources in the radio map in Fig.~\ref{desi_ghost} are embedded within the diffuse emission, showing higher surface brightness compared to the surrounding halo. These sources are indicated in the left panel of Fig.~\ref{desi_ghost} and are consistent with being located at similar redshifts. 
The diffuse emission shows indications of morphological variations. In the north-western region, the radio emission appears to be enhanced around a localised area. A similar feature is observed south of the BCG, where the emission appears to form a boundary-like structure, marked as `radio hole' in Fig.~\ref{desi_ghost} and is discussed in Sect.~\ref{radiohole}. The fact that this feature is not an artefact is discussed in Appendix~\ref{test_artefacts} and shown in Fig.~\ref{fig_test_artefacts}. 

In Fig.~\ref{desi_heart}, a similarly bright central source is observed, coincident with the peak of the diffuse emission and identified as the BCG at a redshift of $z_{\rm photo} = 0.505\pm0.070$. Several galaxies at comparable redshift are distributed across the diffuse radio emission. The extent of this halo is comparable to the previous source, with a size of approximately $\sim430$\,kpc relative to the BCG.

\subsection{Radio spectral analysis}
To investigate the spectral properties of the candidate radio minihalos, we derive both spatially resolved and integrated spectral information based on the MeerKAT L-band data and complementary survey observations. For each minihalo candidate, we construct radial spectral profiles by analysing the emission in concentric annuli matched to the angular resolution of the data. The maps are smoothed to a common resolution of $10\arcsec$, corresponding to the synthesised beam, and circular annuli with widths equal to the beam size are centred on the source. The annuli are successively expanded to trace the emission outwards until the signal approaches the noise level. Within each annulus, we measure the flux density at four representative MeerKAT sub-bands ($1.03$, $1.22$, $1.40$, and $1.52$\,GHz), chosen to provide a wide frequency coverage while maintaining high signal-to-noise ratios. A spectral index is then determined for each radial bin by fitting a power law to the measured flux densities.

In addition to the spatially resolved analysis, we construct integrated radio spectra for the BCG associated with each minihalo candidate. For this purpose, we combine our MeerKAT measurement at 1.28\,GHz with literature data spanning a broad frequency range, including GLEAM-X \citep{gleamx}, GLEAM-300 \citep{Duchesne_2025_gleam300}, RACS-Low \citep{racs}, the ASKAP EMU main survey \citep{emu_main}, as well as RACS-Mid \citep{racsmid} and RACS-High \citep{racshigh}. This allows us to constrain the overall spectral behaviour and identify potential curvature or deviations from a simple power-law spectrum.

Furthermore, we derive pixel-based in-band spectral index maps from the MeerKAT data. To this end, the L-band dataset is split into sub-bands and imaged at a common angular resolution of $10\arcsec$. The spectral index is then determined on a pixel-by-pixel basis by fitting a power law in log\,$S$--log\,$\nu$ space. These maps provide a detailed view of spatial variations in the spectral index across the minihalo candidates and their central galaxies, enabling a direct comparison with the radial profiles and integrated spectra.

The left panel of Fig.~\ref{fig_minihalo_ghost_spectra} shows the radial spectral profile of the minihalo candidate MKT\,J0355--4237. The emission can be traced out to a radius of $\sim300$\,kpc, beyond which it becomes consistent with the noise level. In the central region, the spectral index is approximately $-0.7$. For the BCG, the integrated spectrum, including literature measurements down to 87\,MHz, yields a spectral index of $-0.68 \pm 0.03$. The spectrum indicates a possible change in slope, with steeper values at low frequencies ($87-300$\,MHz) and a flatter behaviour between $\sim800$\,MHz and 1.7\,GHz.
The pixel-based spectral index map shows values of $\sim -0.7$ in the central region, steepening to $\sim-3$ towards lower surface-brightness regions. Similar results are found by \citet{Giacintucci_2014rx}. 

In Fig.~\ref{fig_minihalo_heart_spectra}, the left panel presents the radial spectral profile. The spectral index ranges from $-0.8 \pm 0.3$ in the centre to steeper values of $-1.5 \pm 0.6$ at larger radii. The emission becomes consistent with the noise beyond $\sim300$\,kpc. The integrated spectrum of the central BCG shows a spectral index of $-0.86 \pm 0.02$, with no clear evidence for a change in slope across the observed frequency range.
The pixel-based spectral index map does not reveal significant substructure. It shows spectral index values of approximately $-0.8$ in the central galaxy, steepening to $\sim -2.5$ towards the outskirts. 


\begin{figure*}
    \centering
    \includegraphics[width=\linewidth]{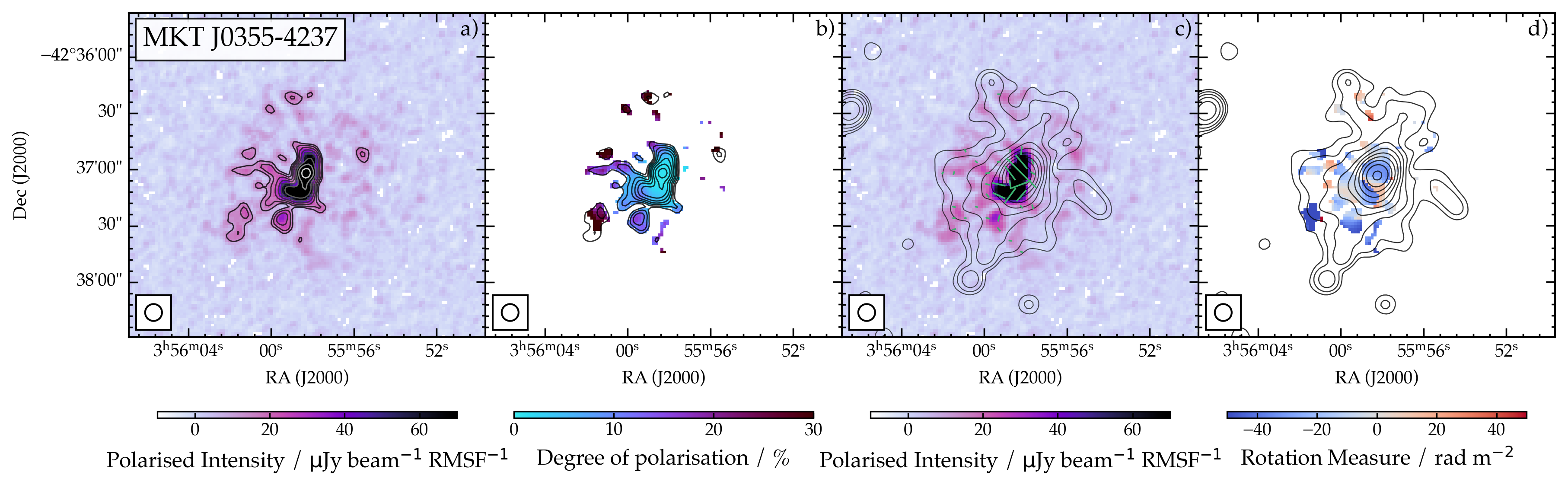}
    \caption{\textit{Panel a):} Polarised intensity of MKT\,J0355--4237 at 1226\,MHz with contours starting at $7\sigma_{\rm PI}$ and increase by $\sqrt{2}$ ($\sigma_{\rm PI}$ =$1.74\,\upmu$Jy\,beam$^{-1}$\,RMSF$^{-1}$).
    \textit{Panel b):} Fractional polarisation at 1226\,MHz with PI contours as in panel a).
    \textit{Panel c):} Polarised intensity with total intensity contours at 1226\,MHz starting at $3\sigma_I$ and increasing by a factor of two ($\sigma_I = 10\,\upmu$Jy/beam). The  magnetic field orientations are shown in black and scale with the polarised intensity.
    \textit{Panel d):} Foreground-corrected RM with total intensity contours as in panel c). For each panel, the circular beam of $9\arcsec$ appears in the lower left corner. The fractional polarisation and the rotation measure maps are clipped to regions within the $3\sigma_I$ contour of the total intensity emission.}
    \label{fig_pol_ghost}
\end{figure*}

\begin{figure*}
    \centering
    \includegraphics[width=\linewidth]{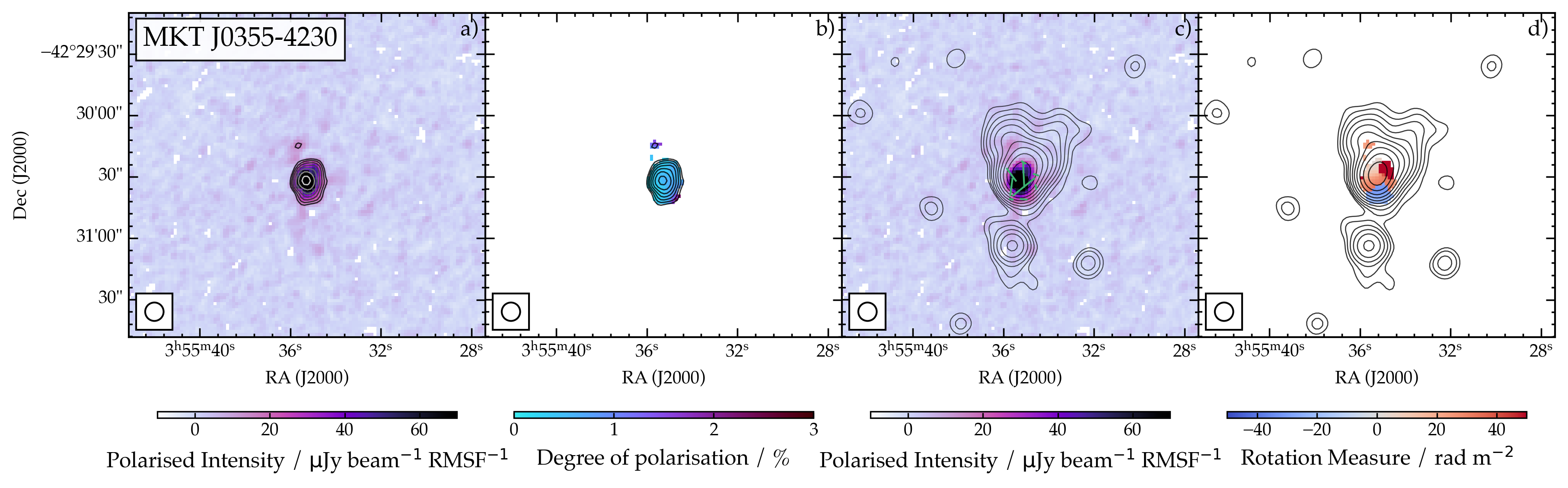}
    \caption{\textit{Panel a):} Polarised intensity of MKT\,J0355--4230 at 1226\,MHz with contours starting at $7\sigma_{\rm PI}$ and increase by $\sqrt{2}$ ($\sigma_{\rm PI}$ =$1.74\,\upmu$Jy\,beam$^{-1}$\,RMSF$^{-1}$).
    \textit{Panel b):} Fractional polarisation at 1226\,MHz with PI contours as in panel a).
    \textit{Panel c):} Polarised intensity with total intensity contours at 1226\,MHz starting at $3\sigma_I$ and increasing by a factor of two ($\sigma_I = 10\,\upmu$Jy/beam). The  magnetic field orientations are shown in black and scale with the polarised intensity.
    \textit{Panel d):} Foreground-corrected RM with total intensity contours as in panel c). For each panel, the circular beam of $9\arcsec$ appears in the lower left corner. }
    \label{fig_pol_heart}
\end{figure*}

\subsection{Polarisation properties}
In Fig.~\ref{fig_pol_ghost} and \ref{fig_pol_heart}, we present the polarised intensity, fractional polarisation, magnetic field orientations\footnote{Throughout this work, `magnetic field orientations’ refers to Faraday rotation corrected magnetic field orientations.}, and foreground-corrected RM maps of the minihalo candidates. To mitigate the bias introduced by the positive-definite noise distribution, we subtract the mean background level (mean$_{\rm PI}=11.51\,\upmu$Jy\,beam$^{-1}$\,RMSF$^{-1}$) from the polarised intensity maps \citep{Heald_2009_II}. We further apply a clipping threshold of $7\,\sigma_{\rm PI}$ to ensure that only significant polarised emission is considered. The noise levels in polarised intensity, $\sigma_{\rm PI}$, are estimated from source-free regions of the bias-corrected maps. As the effective frequency of the polarisation maps derived from RM synthesis corresponds to the weighted mean $\lambda_0^2$, rather than the central observing frequency, the Stokes\,$I$ maps are scaled to a common reference frequency of 1226\,MHz in order to compute the fractional polarisation. To minimise the impact of imaging artefacts in Fig.~\ref{fig_pol_ghost}, the fractional polarisation and rotation measure maps are clipped to regions within the $3\sigma_I$ contour of the total intensity emission.

In Fig.~\ref{fig_pol_ghost}, the polarised intensity distribution shows several notable features. The BCG exhibits the strongest polarised signal, however, the fractional polarisation remains below 1\%. The polarised emission associated with the BCG appears slightly bent from south to north, suggesting a presence of substructure. In contrast, the diffuse halo, particularly in its outer regions, shows lower total polarised intensity but reaches fractional polarisation levels of up to $\sim30$\%. 
It is worth mentioning that such high fractional polarisation in low surface-brightness regions could in principle be attributed to calibration artefacts, which is probably the case here. The Faraday-corrected magnetic field orientations slightly follow the morphology of the polarised emission associated with the BCG. The RM distribution does not exhibit a clear large-scale pattern, which is consistent with a turbulent magnetic field.

There are some pixels with RM values outside the $3\sigma_I$ contours in panel d) of Fig.~\ref{fig_pol_ghost}. These features are smaller than the beam size and may therefore be associated with imaging artefacts. As an additional robustness test, we produced a low resolution image at $22\arcsec$ using \texttt{robust}$ = +0.5$ (described in Appendix~\ref{lowresolution}), which gives more weight to shorter baselines and hence increases the sensitivity to diffuse emission. The resulting morphology, shown in Fig.~\ref{fig:lowresol_polprop} in Appendix~\ref{lowresolution}, is consistent with that of the image in Fig.~\ref{fig_pol_ghost}.  
The few pixels of the RM values outside the $3\sigma_I$ contours are no longer visible in the low resolution image, confirming that the extended emission in the southern region may result from artefact in the imaging. Furthermore, the increased sensitivity to diffuse emission does not reveal a significant contribution from Galactic foreground emission.

\begin{figure*}
    \centering
    \includegraphics[width=0.47\linewidth]{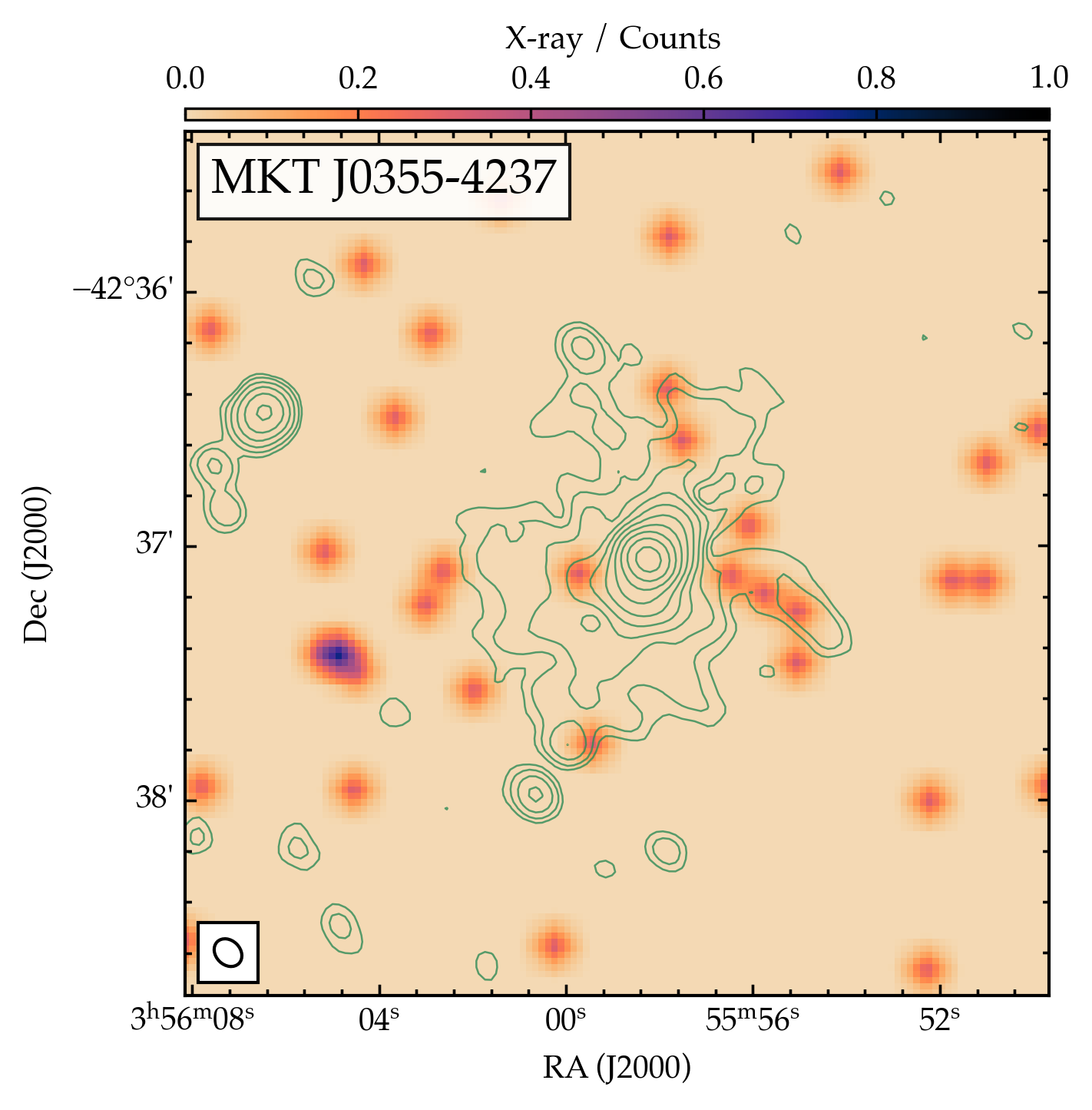}
    \includegraphics[width=0.47\linewidth]{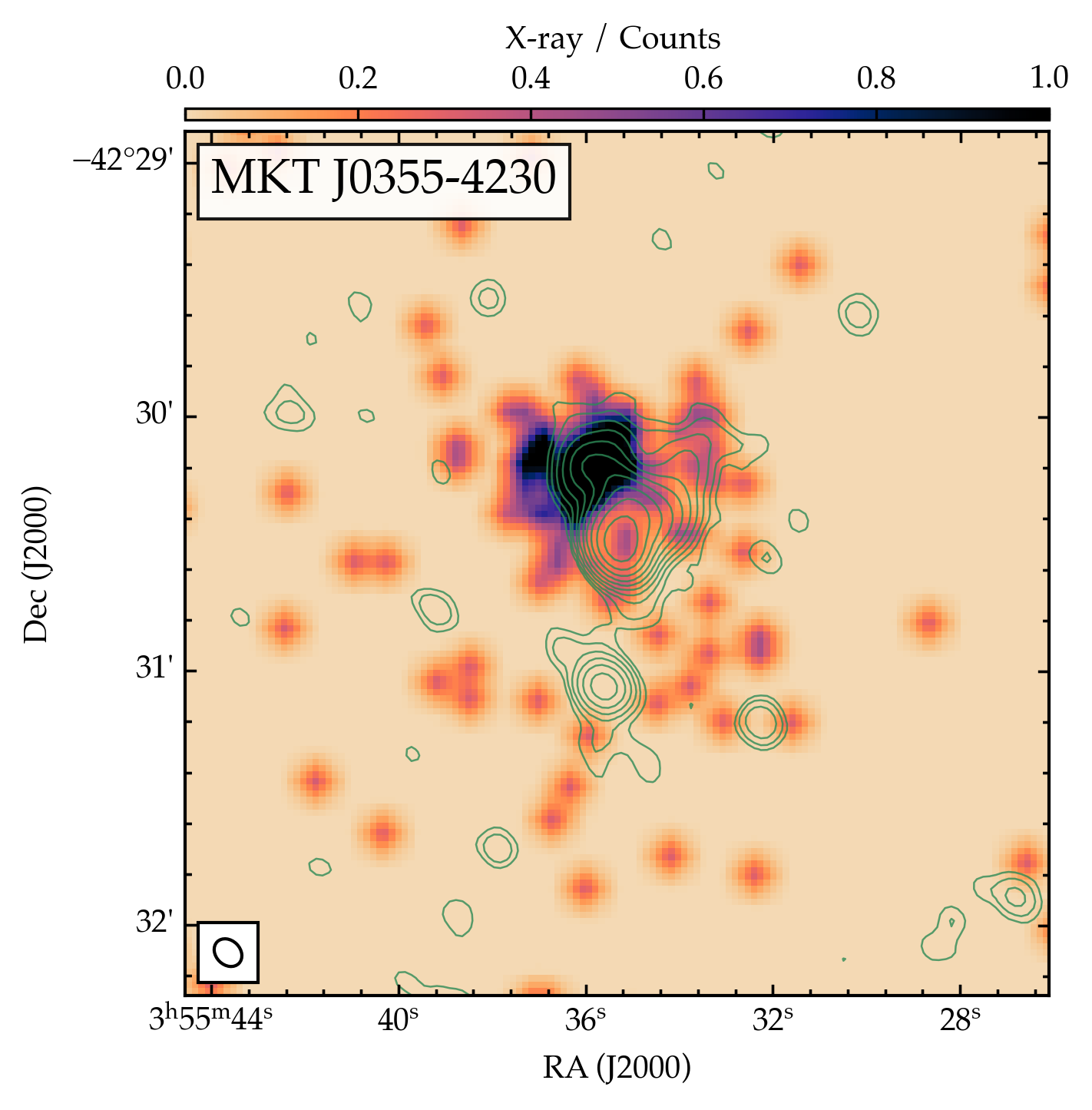}
    \caption{eROSITA X-ray counts map at $0.2-5.0$\,keV of the minihalo candidate smoothed to the Stokes$I$ resolution with overlaid MeerKAT L-band total intensity contours, starting at $3\sigma$ and increasing by factors of 2 ($\sigma=6.5\upmu$Jy/beam). The beam is shown in the bottom left corner.}
    \label{fig_minihalo_xray}
\end{figure*}

In Fig.~\ref{fig_pol_heart}, polarised emission is detected primarily in the central region, coincident with the BCG. Some low-level emission extends towards the north. The fractional polarisation is generally low, remaining below 0.7\% in the central region and increasing to $\sim2.3$\% towards the north. No significant substructure is apparent in the magnetic field orientations, which are broadly aligned along the north–south direction. 
The RM map (panel d) of Fig.~\ref{fig_pol_heart}) shows a change in sign from positive to negative values from north to south. Although this structure is only marginally resolved at the current angular resolution, it may indicate variations in the magnetised ICM. Such variations could arise from turbulent or dynamical processes in the cluster environment, although higher-resolution RM observations are required to characterise the magnetic field structure.

\subsection{X-ray data}
To investigate the thermal properties of the environments hosting the minihalo candidates, we analyse X-ray data, corresponding to the broad energy band of $0.2-5.0$\,keV, obtained with the eROSITA telescope \citep{Predehl_2021}. The data were retrieved from the public eROSITA archive\footnote{\url{https://erosita.mpe.mpg.de}} and consist of photon count maps, which provide a qualitative view of the diffuse X-ray emission. Given the limited depth of the available data, our analysis is restricted to a morphological comparison with the radio emission. The X-ray data of the minihalo candidate, shown in Fig.~\ref{minihalo_envir}, are smoothed to the resolution of the total intensity.

For the source MKT\,J0355--4237, no significant X-ray emission is detected at the location of the radio minihalo candidate, shown in left panel of Fig.~\ref{fig_minihalo_xray}. The X-ray image does not show any clear excess above the background level, either at the position of the BCG or across the extent of the diffuse radio emission. This lack of detectable X-ray emission suggests that either the system is X-ray faint or that the available data are not sufficiently sensitive to trace the underlying ICM.
In contrast,the right panel of Fig.~\ref{fig_minihalo_xray} show the source MKT\,J0355--4230, which exhibits detectable X-ray emission. The emission is not centrally peaked on the BCG, but instead appears enhanced in the northern region, offset from the position of the central radio source. This asymmetry indicates a disturbed or non-uniform distribution of the hot gas. Due to the limited photon statistics, we do not perform a quantitative spectral analysis and instead rely on the count distribution to characterise the morphology of the X-ray emission.
A comparison between the X-ray and radio maps shows that the diffuse radio emission extends beyond the regions of significant X-ray detection. In particular, the displacement between the peak of the X-ray emission and the BCG suggests a possible dynamical interaction within the system, such as sloshing or a recent merger event. However, deeper X-ray observations are required to confirm this scenario and to derive quantitative constraints on the thermodynamical properties of the ICM.


\section{Discussion}
\label{discussion}

\subsection{Origin of the internal radio deficit within the minihalo candidate MKT\,J0355--4237}
\label{radiohole}
In Fig.~\ref{desi_ghost}, we identify a pronounced surface brightness `radio hole' embedded within the diffuse minihalo emission. The surrounding total emission appears to partially enclose this feature, resulting in a well-defined, sharply bounded depression in radio brightness rather than a gradual fading of the emission. In the left panel of Fig.~\ref{minihalo_envir}, we observe that a galaxy at the similar redshift \citep[$z_{\rm photo}=0.36$;][]{Zhou_2023} lies in close proximity to the edge of the deficit region, suggesting a possible association with local dynamical activity or substructure within the cluster potential.
However, the available eROSITA data show no significant X-ray emission at the position of the radio feature, within current sensitivity limits (Fig.~\ref{fig_minihalo_xray}), and no evidence for any excess, cavity, or discontinuity within the source.
This disfavours a strong thermodynamical cavity or shock interpretation, although a shallow or low-contrast ICM perturbation cannot be excluded. In Fig.~\ref{fig_pol_ghost}, the minihalo emission is detected to be polarised on resolved scales at level of less then $1\,\%$. This increases while going to the outskirts of the  diffuse halo. In contrast, no significant polarised emission is detected within the region of the radio deficit. But we emphasise, however, that the angular size of the feature is smaller than the synthesised beam of the polarisation data, and therefore the absence of polarised signal in this region cannot be robustly constrained. Within these limitations, the morphology is nonetheless consistent with a localised suppression of both total and polarised synchrotron emission. 

Similar internal brightness depressions and sharp substructure have been reported in a number of cluster minihalos, particularly in systems \citep[e.g. RX\,J1720.1+2638;][]{Biava_2024,Giacintucci_2014rx} exhibiting sloshing-driven gas motions and cold fronts \citep[e.g.][]{Riseley_2023, VanWeeren_2024}. In these systems, filamentary or patchy radio morphologies are commonly interpreted as the result of spatial variations in turbulence and magnetic field ordering, rather than a uniform relativistic electron distribution. In this context, the observed radio hole may trace a localised region of reduced turbulent re-acceleration efficiency and a change in magnetic field coherence \citep{Gitti_2002}. Shear motions at such interfaces can align and amplify magnetic fields, forming a barrier that confines relativistic electrons and leads to a sharp discontinuity in synchrotron emissivity. The absence of a clear X-ray counterpart does not rule out this scenario, as low-contrast or projection effects may prevent its detection at the current sensitivity. 

An alternative explanation is an AGN-related origin, in which the feature corresponds to a remnant radio lobe within the minihalo. In this case, the observed deficit may arise from aged or displaced synchrotron-emitting plasma. The presence of curved polarised emission across the minihalo suggests locally ordered magnetic fields and may provide indirect support for past or ongoing AGN activity. The WISE colours \citep[W1–W2$ = 0.40 \pm 0.04$ and W2–W3$ = 1.82 \pm 0.24$; ][]{Cutri_2014} are consistent with a passive galaxy or a low-excitation radio AGN, and do not indicate a radiatively efficient nucleus.

To assess whether the observed `radio hole' represents a significant deviation from the intrinsic fluctuations of the diffuse emission, we performed a local surface brightness analysis. We measured the mean surface brightness within a beam-sized aperture centred on the depression and compared it with identical apertures placed around the hole within the surrounding minihalo emission, excluding compact sources and regions outside the diffuse emission. The depression has a mean surface brightness of $6.2\times10^{-5}$\,Jy\,beam$^{-1}$, while the comparison regions have a mean surface brightness of $2.3\times10^{-4}$\,Jy\,beam$^{-1}$ with a standard deviation of $1.7\times10^{-4}$\,Jy\,beam$^{-1}$. Although the depression corresponds to a local minimum in the diffuse emission, its significance is limited to approximately of $1\sigma$ due to the intrinsic brightness variations of the minihalo. We therefore interpret this feature cautiously as a possible surface brightness depression rather than a confirmed radio cavity or deficit. Higher sensitivity and higher-resolution observations will be required to determine whether this structure is associated with a physical discontinuity in the emitting plasma.

\subsection{MKT\,J0355--4230 as a tailed radio galaxy}
Although we classify MKT\,J0355--4230 as a minihalo candidate, its morphology shows similarities to that of a tailed radio galaxy observed at limited angular resolution. Tailed radio galaxies are commonly associated with radio-loud AGN hosted by cluster galaxies, where the interaction between the relativistic plasma launched by the AGN and the surrounding ICM produces elongated, bent, or asymmetric radio structures. The morphology is generally caused by the ram pressure exerted on the radio plasma as the host galaxy moves through the dense cluster environment \citep[e.g.][]{Chen_2020,Giacintucci_2009, Bruno_2024,Merluzzi_2024}. Such sources are preferentially found in galaxy clusters, where the high-density ICM provides the external medium required to bend and confine the radio plasma \citep[e.g.][]{Miley_1980,Giacintucci_2009}.

The diffuse emission of MKT\,J0355--4230 indeed shows an asymmetric morphology that could be interpreted as a low-resolution tailed radio structure. Furthermore, the spectral index distribution shows a steepening towards the outskirts, with values changing from approximately $\alpha\sim-0.8$ in the central region to $\alpha\sim-2.5$ at larger radii. Such spectral steepening is commonly observed in tailed radio galaxies and is generally interpreted as radiative ageing of relativistic electrons as they are transported away from the AGN \citep[e.g.][]{Bruno_2024}. However, similar radial spectral steepening can also occur in diffuse cluster radio sources due to the ageing and spatial variation of the electron population.
Tailed radio galaxies often show significant fractional polarisation, particularly along the tails, where shear and compression of the plasma can order the magnetic field \citep[e.g.][]{Feretti_1998,Wezgowiec_2024}. In contrast, MKT\,J0355--4230 shows very low fractional polarisation in the central region, below $0.7\%$, increasing only towards the northern outskirts up to $\sim2.3\%$. The observed polarisation is therefore lower than typically reported for well-resolved tailed radio galaxies, although beam depolarisation and limited angular resolution may reduce the measured polarised fraction.

A further test is to investigate the diffuse emission after removing the short baselines from the data during imaging. The full procedure is described in Appendix~\ref{uvcut}. In Fig.~\ref{fig:uvcut}, we show MKT\,J0355--4230 imaged after applying an $uv$cut of $20$\,k$\lambda$ and $26$\,k$\lambda$, respectively. Negative residuals are observed around the BCG in both images, demonstrating that a substantial fraction of the emission is on spatial scales sampled by the short baselines and is therefore associated with diffuse, extended emission. The integrated flux density decreases from $50.3$\,mJy in the image without a $uv$cut to $17.0$ and $8.2$\,mJy for the $20$\,k$\lambda$ and $26$\,k$\lambda$ cuts, respectively. This substantial decrease indicates that a significant fraction of the emission is associated with spatially extended structures. The available angular resolution limits our ability to further characterise the morphology and nature of this diffuse emission.

\begin{table*}
\renewcommand{\arraystretch}{1.3}
\centering
\caption{RM and magnetic field properties of the two minihalo candidates.}
\begin{tabular}{lcc}
\toprule
Parameter & MKT\,J0355--4237 & MKT\,J0355--4230 \\
\midrule
$\langle \mathrm{RM} \rangle$ / rad m$^{-2}$ & $-14.41 \pm 10.30$ & $15.47 \pm 31.84$ \\
$\sigma_{\mathrm{RM}}$ / rad m$^{-2}$ & $57.13 \pm 7.41$ & $90.70 \pm 24.05$ \\
$B_{\mathrm{ord}}$ / $\upmu$G & $-0.134^{+0.100}_{-0.182}$ & $0.126^{+0.374}_{-0.272}$ \\
$B_{\rm turb}$ / $\upmu$G & $1.173^{+0.830}_{-0.488}$ & $1.264^{+0.999}_{-0.570}$ \\
\bottomrule
\end{tabular}
\label{rmanalysis_prop_table}
\end{table*}

\subsection{Disentangling turbulent and ordered magnetic field}
\label{faraday_analysis_discussion}
Building on the RM maps presented in Sect.~\ref{fig_pol_ghost} and \ref{fig_pol_heart}, we use the Faraday rotation signal to quantify the magneto-ionic properties of the ICM. The RM provides a direct probe of the line-of-sight integral of the thermal electron density and magnetic field, and can therefore be used to constrain both the ordered and turbulent components of the cluster magnetic field \citep{Brentjens_2005, Beck_2019, Takahashi_2023}.
Under the assumption that the Faraday rotation arises predominantly within the ICM, the RM is given by
\begin{equation}
    \mathrm{RM} = 0.812 \int n_{\rm e}\, B_\parallel \, dl \quad,
\end{equation}
where $n_{\rm e}$ is the electron density in ${\rm cm^{-3}}$, $B_\parallel$ is the magnetic field component along the line of sight in $\upmu$G, and $l$ is the path length in pc. The analysis is restricted to a circular region centred on the minihalo. Only pixels of the RM within the $3\sigma$ region of the Stokes~$I$ image are taken into the mask. Within this region, we compute the mean RM $\langle {\rm RM} \rangle$ and its dispersion $\sigma_{\rm RM}$. The latter is corrected for measurement uncertainties. The $\langle {\rm RM} \rangle$ traces the ordered component of the magnetic field along the line of sight, while the $\sigma_{\rm RM}$ reflects fluctuations associated with a turbulent field\footnote{Following the notation of \citet{Beck_2015}, we assume that the ordered magnetic field is the regular magnetic field $B_{\rm ord}=B_{\rm reg}$ and the turbulent magnetic field is the isotropic magnetic field $B_{\rm turb}=B_{\rm iso}$}. Assuming a uniform electron density $n_{\rm e}$ of $0.0003\,$cm$^{-3}$ in a cluster environment and a path length $L$ of the size of the minihalos (see Table~\ref{rmanalysis_prop_table}), the magnetic field strengths are estimated as
\begin{equation}
    B_{\mathrm{ord}} = \frac{\langle \mathrm{RM} \rangle}{0.812\, n_e\, L}
\end{equation}
and 
\begin{equation}
    B_{\mathrm{turb}} = \frac{\sigma_{\mathrm{RM}}}{0.812\, n_{\rm e}\, \sqrt{L\, l_0}},
\end{equation}
where $l_0$ denotes the magnetic field coherence scale and is set to 50\,kpc \citep{Beck_2015,Vacca_2018}. Here we assume a fiducial value of $l_0$ based on commonly used values in the literature. A through estimate of $l_0$ is possible via analysing the structure function and/or power spectrum of RM \citep{Stil_2011, Seta_2024}.
However, the limited range of spatial scales covered by our RM maps does not allow a reliable determination of $l_0$ from these methods for the present data. A future work will deal these methods.

Uncertainties are estimated from the finite number of independent resolution elements, defined as the ratio between the analysed area and the effective beam area. These statistical uncertainties are propagated via Markov Chain Monte Carlo sampling of $\langle \mathrm{RM} \rangle$, $\sigma_{\rm RM}$, and the physical parameters $n_{\rm e}$, $L$, and $l_0$, which are assumed to follow log-normal distributions. The magnetic field coherence scale $l_0$ is additionally constrained to remain larger than the beam size, ensuring that only physically unresolved (and not beam-smeared) turbulence scales are considered.

For the minihalo MKT\,J0355--4237, we measure a modest mean RM of $\langle \mathrm{RM} \rangle = -14.4 \pm 10.3\,\mathrm{rad\,m^{-2}}$, consistent with a weak large scale coherent magnetic field along the line of sight. 
In contrast, the RM dispersion of $\sigma_{\mathrm{RM}} = 57.1 \pm 7.4\,\mathrm{rad\,m^{-2}}$, indicating significant small-scale fluctuations. The inferred magnetic field strengths are $B_{\rm ord} \sim 0.13^{+0.10}_{-0.18}\,\upmu$G and $B_{\rm turb} \sim 1.17^{+0.87}_{-0.48}\,\upmu$G, implying that the magnetic field is dominated by its turbulent component. This is consistent with the absence of a clear large-scale RM pattern and supports a scenario in which the ICM is permeated by a tangled magnetic field. 

A similar result was obtained for the minihalo candidate MKT\,J0355--4230, although with larger uncertainties due to the smaller number of independent RM measurements. The mean RM, $\langle \mathrm{RM} \rangle = 15.5 \pm 31.8\,\mathrm{rad\,m^{-2}}$, is consistent with zero within the uncertainties, again indicating no strong coherent line-of-sight field. However, the RM dispersion is higher, $\sigma_{\mathrm{RM}} = 90.7 \pm 24.1\,\mathrm{rad\,m^{-2}}$, suggesting enhanced magnetic field fluctuations or a longer effective path length through the magnetised medium. The corresponding magnetic field estimates, $B_{\rm ord} \sim 0.14^{+0.37}_{-0.27}\,\upmu$G and $B_{\rm turb} \sim 1.27^{+0.88}_{-0.57}\,\upmu$G, reinforce the dominance of turbulent magnetic fields. These values should be considered as estimates and we leave the more detailed analysis using the structure function and power spectrum of the RM to future observations with improved spatial resolution.

\section{Conclusion}
\label{conclusion}
We present the discovery and analysis of two diffuse radio sources, MKT\,J0355--4237, identified as a candidate minihalo, and MKT\,J0355--4230, whose nature remains still uncertain between a minihalo candidate and a tailed radio galaxy.
Both systems exhibit extended ($\sim$400–430 kpc) low-surface-brightness synchrotron emission centred on the brightest cluster galaxy, confirming their nature as cluster-scale non-thermal phenomena in dynamically active environments. Analysing these systems, we come the following conclusions
\begin{enumerate}
   \item The radio emission in both sources exhibits spatial variations in spectral index. In MKT\,J0355--4237, the flattest spectral indices ($\alpha \approx -0.7$ to $-0.8$) are associated with the central BCG/AGN region, while the surrounding diffuse emission remains comparatively uniform, with only moderate variations on scales comparable to the beam size. In MKT\,J0355--4230, the spectral index appears to steepen preferentially toward the northern regions, which may indicate synchrotron ageing of relativistic electrons, although the current data do not allow a definitive interpretation. Overall, the observed spectral behaviour is broadly consistent with diffuse cluster-related radio emission, while also allowing for possible contributions from aged AGN plasma.

    \item Polarised emission is detected in both minihalos, with low central fractional levels ($<1\,\%$) that increase toward larger radii, reaching up to $\sim$30\,\% in MKT\,J0355--4237. This indicates that, despite the turbulent environment, the intracluster magnetic field retains a degree of local ordering. In MKT\,J0355--4230, the polarisation and RM structure further suggest coherent large-scale magnetic field orientation, including a possible large-scale sign change in Faraday depth.

    \item The two systems also differ in their thermodynamical and dynamical states. MKT\,J0355--4230 is associated with a previously uncatalogued galaxy cluster identified via galaxy overdensity techniques, and shows an X-ray morphology offset from the BCG, indicative of dynamical activity such as gas sloshing or a recent merger. In contrast, MKT\,J0355--4237 is embedded in a known cluster and displays more centrally concentrated X-ray emission, but contains pronounced internal radio substructure.

    \item MKT\,J0355--4237 additionally exhibits a sharply bounded internal surface brightness depression, like a `radio hole', embedded within the diffuse emission. No corresponding X-ray cavity or shock is detected in available eROSITA data. This feature may therefore trace a localised region of reduced turbulent re-acceleration within the minihalo emission.

    \item MKT\,J0355--4230 shows a morphology and spectral-index gradient consistent with a tailed radio galaxy, with $\alpha$ steepening from approximately $-0.8$ in the centre to $-2.5$ towards the outskirts. However, its very low fractional polarisation ($<0.7\,\%$, increasing to $\sim2.3\,\%$ in the northern outskirts) is lower than typically observed in well-resolved tailed radio galaxies. Thus, while a tailed-radio-galaxy interpretation is plausible, the available data quality do not allow us to clearly classify MKT\,J0355--4230 as either a true minihalo or a tailed radio galaxy.

    \item The Faraday rotation analysis indicates that both minihalos are dominated by turbulent magnetic fields at the $\sim 1\,\upmu$G level, while the ordered component is weak and consistent with zero within uncertainties. 

\end{enumerate}

\bibliographystyle{aa} 
\bibliography{literatur} 

\begin{acknowledgements}
ST, CJR and DJB acknowledge the support from the DFG via the Collaborative Research Center SFB1491 \textit{Cosmic Interacting Matters - From Source to Signal}.
AB gratefully acknowledge the financial support provided by the German Federal Ministry of Research, Technology and Space (BMFTR) in the framework of the Knowledge creates perspectives for the region!, for the project StStG – DZA – Aufbauphase: Deutsches Zentrum für Astrophysik, Großforschungszentrum in der sächsischen Lausitz: Aufbauphase 2026, grant number 03WSP1746.
The MeerKAT telescope is operated by the South African Radio Astronomy Observatory, which is a facility of the National Research Foundation, an agency of the Department of Science and Innovation.
This paper includes archived data obtained through the CSIRO ASKAP Science Data Archive, CASDA (http://data.csiro.au).
We acknowledge data storage and computational facilities by the University of Bielefeld which are hosted by the Forschungszentrum Jülich and that were funded by German Federal Ministry of Education and Research (BMBF) projects D-LOFAR IV (05A17PBA) and D-MeerKAT-II (05A20PBA), as well as technical and operational support by BMBF projects D-LOFAR 2.0 (05A20PB1) and D-LOFAR-ERIC (05A23PB1), and German Research Foundation (DFG) project PUNCH4NFDI (460248186). 
The Photometric Redshifts for the Legacy Surveys (PRLS) catalog used in this paper was produced thanks to funding from the U.S. Department of Energy Office of Science, Office of High Energy Physics via grant DE-SC0007914.
The Legacy Surveys consist of three individual and complementary projects: the Dark Energy Camera Legacy Survey (DECaLS; Proposal ID \#2014B-0404; PIs: David Schlegel and Arjun Dey), the Beijing-Arizona Sky Survey (BASS; NOAO Prop. ID \#2015A-0801; PIs: Zhou Xu and Xiaohui Fan), and the Mayall z-band Legacy Survey (MzLS; Prop. ID \#2016A-0453; PI: Arjun Dey). DECaLS, BASS and MzLS together include data obtained, respectively, at the Blanco telescope, Cerro Tololo Inter-American Observatory, NSF’s NOIRLab; the Bok telescope, Steward Observatory, University of Arizona; and the Mayall telescope, Kitt Peak National Observatory, NOIRLab. Pipeline processing and analyses of the data were supported by NOIRLab and the Lawrence Berkeley National Laboratory (LBNL). The Legacy Surveys project is honored to be permitted to conduct astronomical research on Iolkam Du’ag (Kitt Peak), a mountain with particular significance to the Tohono O’odham Nation.
The Legacy Survey team makes use of data products from the Near-Earth Object Wide-field Infrared Survey Explorer (NEOWISE), which is a project of the Jet Propulsion Laboratory/California Institute of Technology. NEOWISE is funded by the National Aeronautics and Space Administration.
The Legacy Surveys imaging of the DESI footprint is supported by the Director, Office of Science, Office of High Energy Physics of the U.S. Department of Energy under Contract No. DE-AC02-05CH1123, by the National Energy Research Scientific Computing Center, a DOE Office of Science User Facility under the same contract; and by the U.S. National Science Foundation, Division of Astronomical Sciences under Contract No. AST-0950945 to NOAO.
This work is based on data from eROSITA, the soft X-ray instrument aboard SRG, a joint Russian-German science mission supported by the Russian Space Agency (Roskosmos), in the interests of the Russian Academy of Sciences represented by its Space Research Institute (IKI), and the Deutsches Zentrum für Luft- und Raumfahrt (DLR). The SRG spacecraft was built by Lavochkin Association (NPOL) and its subcontractors, and is operated by NPOL with support from the Max Planck Institute for Extraterrestrial Physics (MPE). The development and construction of the eROSITA X-ray instrument was led by MPE, with contributions from the Dr. Karl Remeis Observatory Bamberg \& ECAP (FAU Erlangen-Nuernberg), the University of Hamburg Observatory, the Leibniz Institute for Astrophysics Potsdam (AIP), and the Institute for Astronomy and Astrophysics of the University of Tübingen, with the support of DLR and the Max Planck Society. The Argelander Institute for Astronomy of the University of Bonn and the Ludwig Maximilians Universität Munich also participated in the science preparation for eROSITA.
\end{acknowledgements}

\begin{appendix}
\onecolumn

\section{Imaging robustness test}
\label{test_artefacts}
To assess whether the radio hole' is introduced by the imaging procedure, we re-imaged the data using Briggs robust weighting parameters spanning \texttt{robust}$=-2$ to \texttt{robust}$=0.5$, ranging from uniform to near-natural weighting. We could not adopt higher robust values because the images reached the confusion limit at \texttt{robust}$=0.5$. We find that the depression is consistently recovered across all weighting schemes that provide a reasonable compromise between sensitivity and angular resolution (\texttt{robust} $\lesssim 0.5$), with the morphology remaining stable and the radio hole’ clearly visible. At higher robust values, the feature becomes progressively less distinct and eventually disappears, consistent with the reduced angular resolution of naturally weighted images. The persistence of the morphology across a broad range of weighting schemes therefore indicates that the `radio hole’ is unlikely to be an artefact of a particular imaging configuration.

\begin{figure*}[h]
    \centering
    \includegraphics[width=\linewidth]{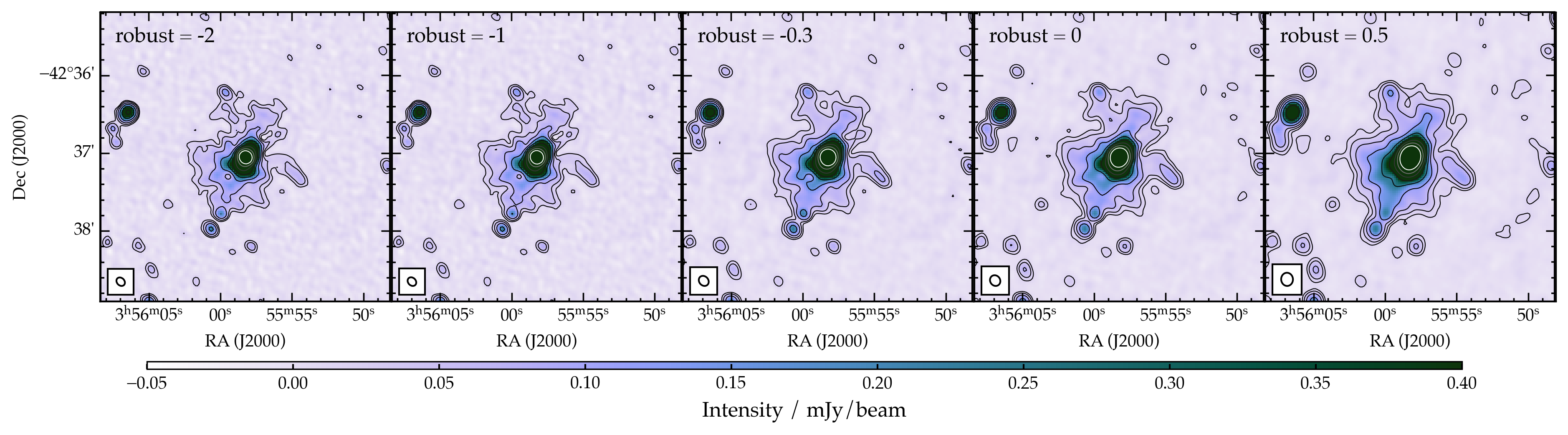}
    \caption{Total intensity of MKT\,J0355--4237 using different Briggs robust weightings, ranging from $-2$ to $0.5$. The contours start at $3\sigma$ and increasing by a factor of 2 at a central frequency of 1.28\,GHz with a noise level $\sigma$ of $6.5, 6.3, 5.5, 5, 5\,\upmu$Jy/beam (from left to right). The beam is shown in the bottom left corner.}
    \label{fig_test_artefacts}    
\end{figure*}

\section{Low-resolution polarisation properties of MKT\,J0355--4237}
\label{lowresolution}
\begin{figure*}[h]
    \centering
    \includegraphics[width=\linewidth]{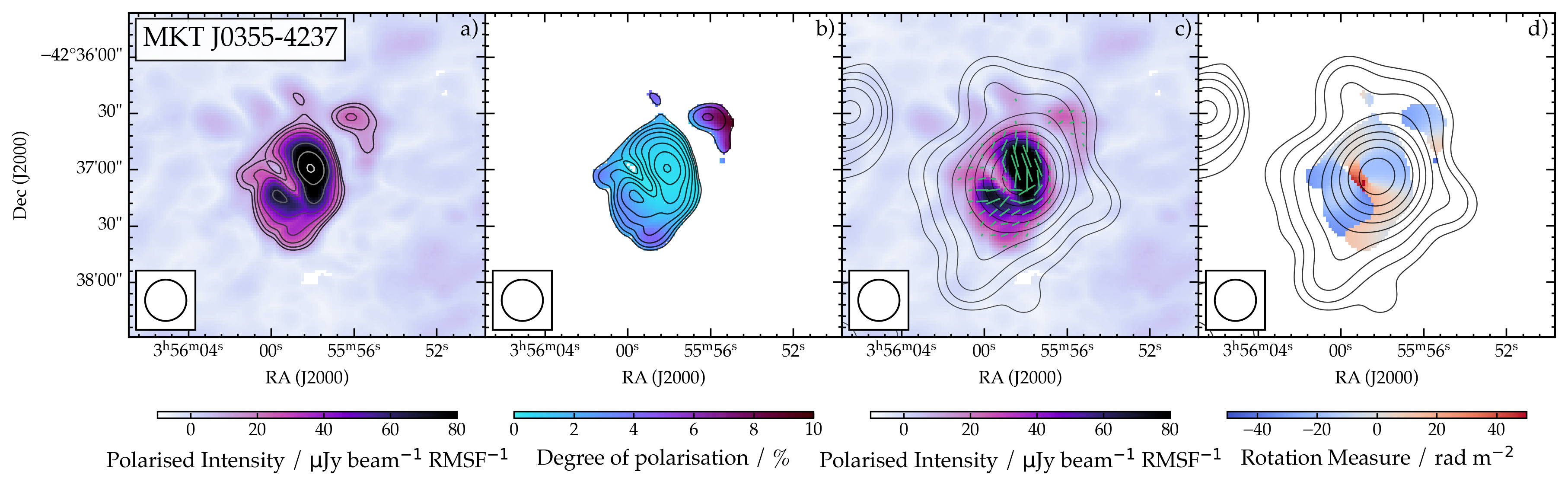}
    \caption{\textit{Panel a):} Polarised intensity of MKT\,J0355--4237 at 1226\,MHz with contours starting at $7\sigma_{\rm PI}$ and increase by $\sqrt{2}$ ($\sigma_{\rm PI}$ =$1.64\,\upmu$Jy\,beam$^{-1}$\,RMSF$^{-1}$).
    \textit{Panel b):} Fractional polarisation at 1226\,MHz with PI contours as in panel a).
    \textit{Panel c):} Polarised intensity with total intensity contours at 1226\,MHz starting at $3\sigma$ and increasing by a factor of two ($\sigma = 20\,\upmu$Jy/beam). The  magnetic field orientations are shown in black and scale with the polarised intensity.
    \textit{Panel d):} Foreground-corrected RM with total intensity contours as in panel c). For each panel, the circular beam of $22\arcsec$ appears in the lower left corner.}
    \label{fig:lowresol_polprop}
\end{figure*}

To test the robustness of the polarised emission of MKT\,J0355--4237, we repeat the polarisation imaging described in Sect.~\ref{datameerkat} using a \texttt{robust} parameter of $+0.5$, resulting in a lower angular resolution of $22\arcsec$ and increased sensitivity to emission on larger spatial scales. We perform RM synthesis and RM-Clean following the same procedure as for the standard image and correct the polarised intensity for Rician bias. We additionally subtract a mean background level of $14.2\,\upmu$Jy\,beam$^{-1}$\,RMSF$^{-1}$, as determined from the off-source regions. Fig.~\ref{fig:lowresol_polprop} shows the resulting polarised intensity, degree of polarisation, Faraday-corrected magnetic-field orientation, and Galactic-foreground-corrected RM map. The low-resolution image recovers the same overall polarised structure as the fiducial image, while the small-scale features seen at higher resolution are largely smoothed out. This supports their interpretation as small-scale imaging artefacts rather than extended polarised emission. Furthermore, no additional large-scale emission associated with the Galactic foreground is revealed, arguing against a significant contribution from Galactic foreground emission.

\section{Description of the $uv$cut experiment on MKT\,J0355--4230}
\label{uvcut}
\begin{figure*}[h]
    \centering
    \includegraphics[width=0.8\linewidth]{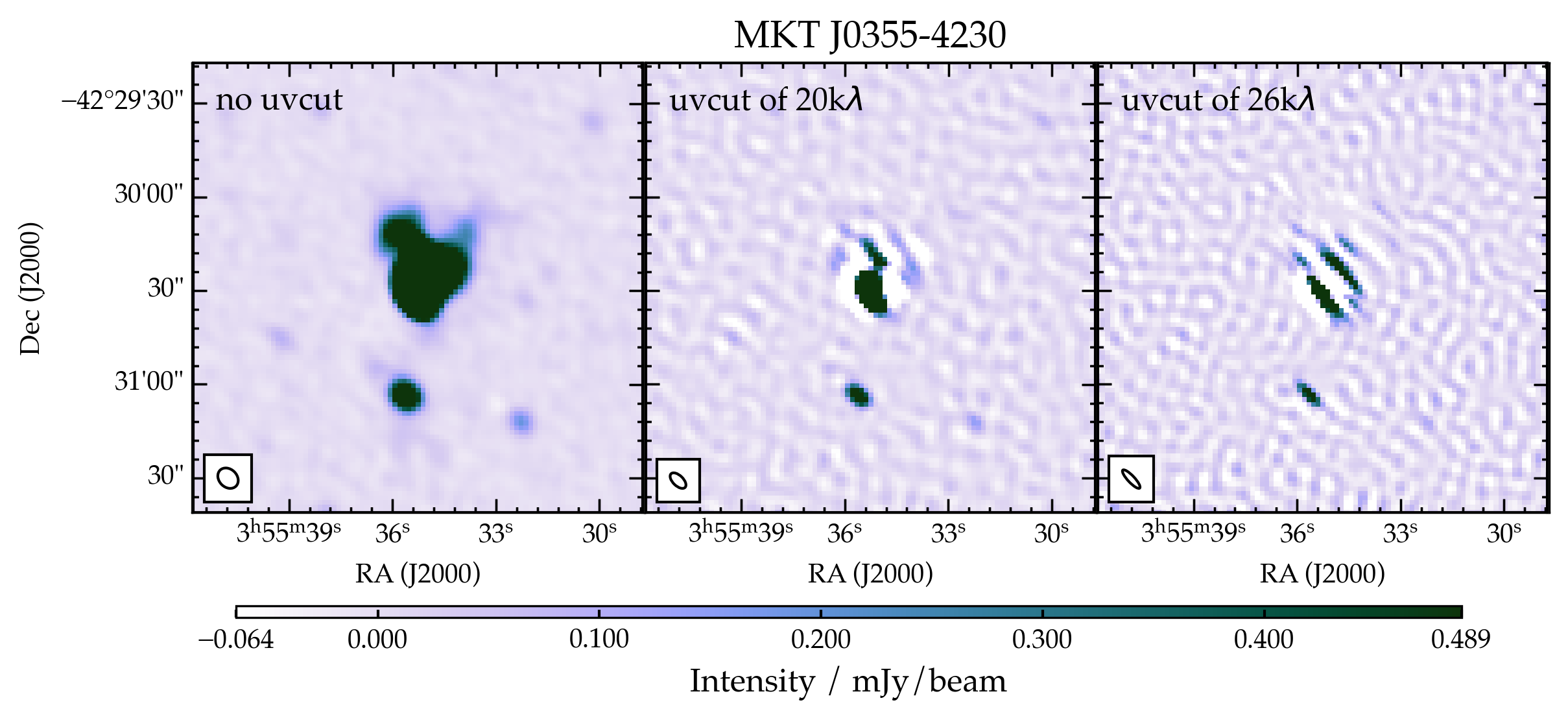}
    \caption{Stokes $I$ images of MKT\,J0355--4230 obtained using the full $uv$ range (left), and after applying minimum $uv$-distance cuts of $20$\,k$\lambda$ (middle) and $26$\,k$\lambda$ (right). The restoring beam is shown in the lower-left corner of each panel.}
    \label{fig:uvcut}
\end{figure*}

To test whether the diffuse emission belongs to the nature of a minihalo or if it contains compact substructures, such as jets or lobes, we repeated the imaging while progressively removing the shortest baselines. This suppresses sensitivity to large spatial scales and allows us to assess whether emission on smaller scales remains detectable. 
We followed the imaging procedure described in Sect.~\ref{datameerkat}, but additionally applied a minimum $uv$-distance using the \textsc{WSClean} parameter \texttt{-minuv-l}. We considered an $uv$cut of $20$\,k$\lambda$ and $26$\,k$\lambda$, thereby retaining only the longer baselines. The resulting images are shown in Fig.~\ref{fig:uvcut} together with the image obtained using the full $uv$ range.

\end{appendix}

\end{document}